\makeatletter
\declare@file@substitution{revtex4-1.cls}{revtex4-2.cls}
\makeatother

\documentclass[10pt, preprint]{aastex631}

\usepackage{graphicx}
\usepackage{subfigure}
\usepackage{subfloat}
\usepackage{amssymb}
\usepackage{amsmath}
\usepackage{multirow}
\usepackage{dcolumn}
\usepackage{bm}
\usepackage{wasysym}
\usepackage{cancel}
\usepackage{makecell}
\usepackage{float}

\shorttitle{$\alpha$ and $\beta$ Effects in a Rotating Astroplasma Sphere}
\shortauthors{}

\begin{document}

\title{Magnetic Field Amplification and Reconstruction in Rotating Astrophysical Plasmas: Verifying the Roles of $\alpha$ and $\beta$ in Dynamo Action}

\author{Kiwan Park}
\affiliation{Institute of Plasmas Turbulence and Magnetic Fields (IPTM) and Soonsil University,\\369 Sangdo-ro Dongjak-gu Seoul 06978 Republic of Korea\\ pkiwan@ssu.ac.kr, pkiwan@gmail.com\\}

\date{\today}

\begin{abstract}
We investigated the $\alpha$ and $\beta$ effects in a rotating spherical plasma system relevant to astrophysical environments. These coefficients were derived using three different approaches based on the large-scale magnetic field $\overline{\mathbf{B}}$, turbulent velocity $\mathbf{u}$, and turbulent magnetic field $\mathbf{b}$, yielding $\alpha_{\mathrm{EM-HM}}$, $\beta_{\mathrm{EM-HM}}$, $\beta_{\mathrm{vv-vw}}$, and $\beta_{\mathrm{bb+jb}}$. Using raw data from direct numerical simulations (DNS), we constructed the magnetic induction equation incorporating the $\alpha$ and $\beta$ coefficients. We then reproduced the $\overline{\mathbf{B}}$ field and compared the results with the DNS data. In the kinematic regime, where $\overline{\mathbf{B}}$ is weak, all models exhibit good agreement with the DNS results. However, in the nonlinear regime, the $\overline{\mathbf{B}}$ field, reproduced using $\beta_{\mathrm{vv-vw}}$, deviates from the DNS and exhibits unbounded growth. To address this discrepancy, we added $\beta_{\mathrm{bb+jb}}$, which represents the contribution of turbulent magnetic fields, to $\beta_{\mathrm{vv-vw}}$. This addition suppresses the divergent growth of $\overline{\mathbf{B}}$ in the nonlinear regime. We then assessed the actual influence of $\alpha$ and $\beta$ on the evolution of $\overline{\mathbf{B}}$ by applying weighted combinations of the two coefficients. Our results show that magnetic $\beta$ diffusion plays a dominant role throughout the entire process. In contrast, the $\alpha$ effect is minor in the kinematic regime but becomes essential for sustaining the $\overline{\mathbf{B}}$ field in the nonlinear regime. We also discussed the underlying physical mechanism responsible for this behavior.
\end{abstract}


\section{Introduction}

\begin{figure*}
  \centering
  \includegraphics[width= 20 cm]{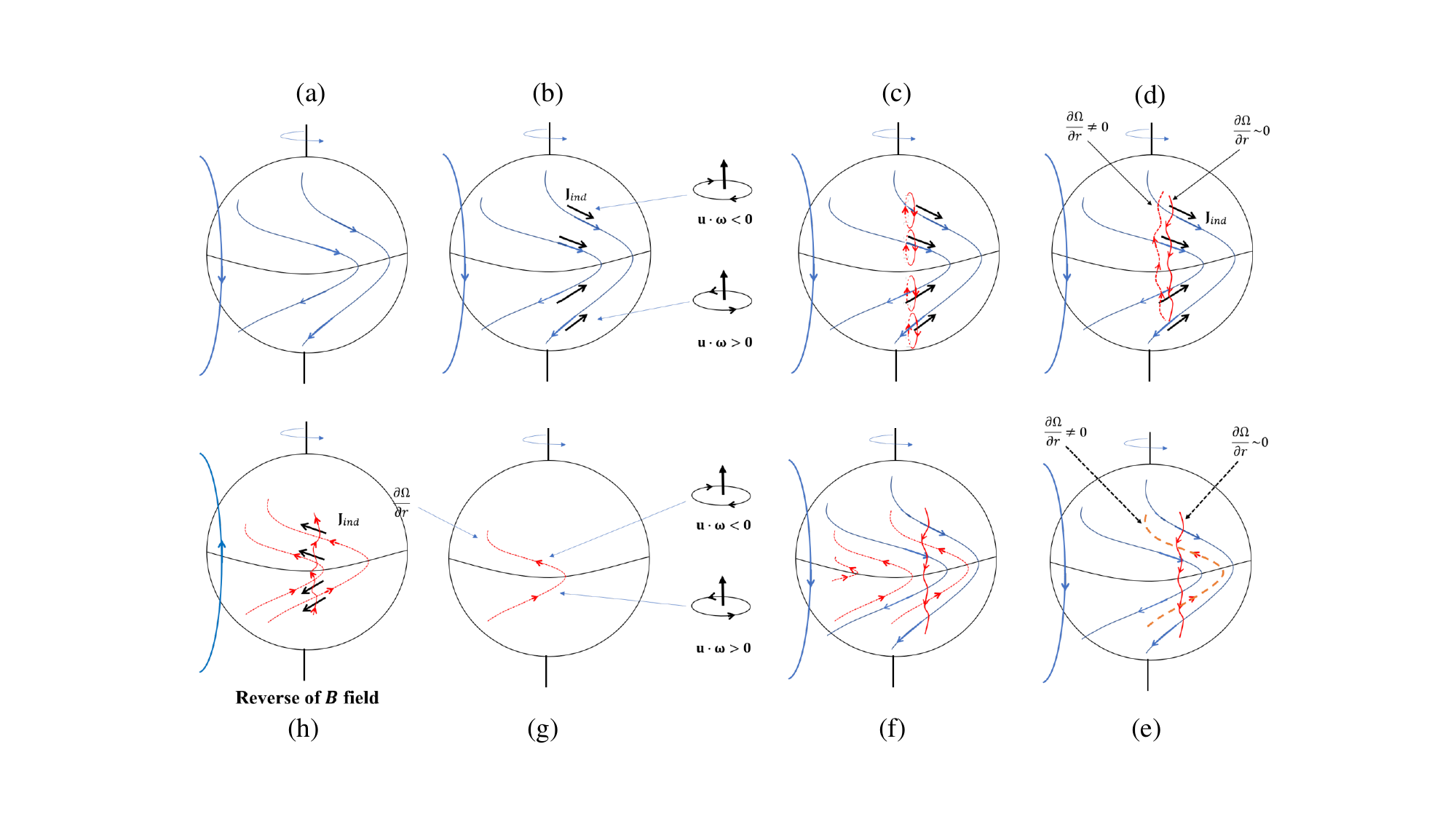}
  \label{f1}
  \caption{Evolution of magnetic fields in the northern and southern hemisphere. Note the directions of kinetic helicity and current density in each hemisphere.}
\end{figure*}

Rotating plasma structures, such as stars, accretion disks, and other similar systems, make unique physical properties, including buoyancy and Coriolis forces. These phenomena give rise to kinetic helicity, defined as $\langle\mathbf{U} \cdot \nabla \times \mathbf{U}\rangle$, where $\mathbf{U}$ represents the fluid velocity. Kinetic helicity, in turn, generates conserved magnetic helicity $\langle\mathbf{A} \cdot \mathbf{B}\rangle$, where $\mathbf{B} = \nabla \times \mathbf{A}$. These pseudo-scalars, along with kinetic and magnetic energy, contribute to the $\alpha$ and $\beta$ effects, which govern the evolution of magnetic fields in helical plasmas.\\

A representative example of the magnetic field generated in the rotating spherical plasma system is the evolution of the solar magnetic field. The Sun's 22-year magnetic cycle, known as the Hale cycle, involves the strengthening, weakening, and polarity reversal of its magnetic field, reflecting internal activity and contributing to its long-term stability. The evolution of the solar magnetic field can be linearly described by the $\alpha$ and $\beta$ effects. These coefficients not only serve as a linear framework for modeling magnetic field evolution but also provide highly effective theoretical tools for analyzing the interaction between magnetic fields and plasma \citep{1999GMS...111..301B}. Several models have been proposed to calculate these quantities analytically or numerically.  The $\alpha$ effect is known to amplify the large-scale magnetic field and determine the polarity of magnetic helicity. It also couples the toroidal magnetic flux, $\mathbf{B}_{\mathrm{tor}}$, and the poloidal magnetic flux, $\mathbf{B}_{\mathrm{pol}}$, leading to their periodic evolution \citep{1955ApJ...122..293P, 1975JFM....68..769F, 1978mfge.book.....M, 1980opp..bookR....K, 2001ApJ...550..824B, 2005AN....326..245S, Akira2011}. In contrast, the $\beta$ effect is associated with magnetic field diffusion, but its influence extends beyond mere magnetic dissipation \citep{2020ApJ...898..112P, 2025PhRvD.111b3021P}.\\

The first conceptual model for the $\alpha$ effect was proposed by \cite{1955ApJ...122..293P}. The author suggested that buoyancy lift the toroidal magnetic flux, $\mathbf{B}_{\text{tor}}$, and that the Coriolis force twist the rising magnetic tube by an angle of approximately $\pi/2$ through the $\alpha$ effect, resulting in a rotated magnetic loop and the generation of poloidal magnetic flux, $\mathbf{B}_{\text{pol}}$. The twisted angle is given by $\theta = 2\omega H/v$, where $\omega$ is the angular velocity, $H$ is the vertical displacement (height), and $v$ is the rising speed. Subsequently, the differential rotation (the $\Omega$-effect) or shear converts $\mathbf{B}_{\text{pol}}$ back into $\mathbf{B}_{\text{tor}}$, thereby completing the cycle. However, the terminology and symbolic representation of the $\alpha$ effect are not equivalent to those in modern dynamo theory. In particular, the explicit twist angle is not a parameter considered in the formulation of the $\alpha$ effect.\\

The $\alpha$ effect acts in conjunction with the current density $\partial \mathbf{B}/\partial t \sim \alpha \mathbf{J}$(current density). The buoyancy and Coriolis forces give rise to kinetic helicity $\langle \mathbf{U} \cdot \nabla \times \mathbf{U} \rangle$, and then this helicity generates current density $\mathbf{J}$ with the seed magnetic field, which in turn, produces the helical magnetic structure. Theoretically, this process is captured in the curl of the electromotive force (EMF, $\mathbf{U} \times \mathbf{B}$). It should be noted that the EMF is not associated with a mechanical force acting on a plasma flux tube, but an effective electric force that drives charged particles, thereby generating $\mathbf{J}(\sim \mathbf{E}+\mathbf{V}\times \mathbf{B}$).\\


On the other hand, the Babcock-Leighton (BL, \citep{1961ApJ...133..572B}) dynamo model is characterized by the transport of magnetic flux through single or multiple meridional circulation cells, along with other transport processes such as directional turbulent pumping and isotropic diffusive transport. The BL model has been successful in simulating some aspects of the solar cycle, particularly the regeneration of poloidal magnetic fields from decaying sunspots, but it also faces limitations \citep{2007AdSpR..39.1661C}. Whether the meridional circulation and other transport processes exist as assumed remains unclear, as they depend on internal motions presupposing an inhomogeneous distribution of density and temperature in an (almost) axisymmetric system. Furthermore, the timescales and patterns of meridional circulation inferred from helioseismology are not fully consistent with the predictions of the BL model, raising questions about the robustness of the assumed flows.\\

These two models essentially include the $\alpha$ and $\beta$ effects along with rotational effects. However, as pointed out earlier, the $\alpha$ effect in the Parker model, the Babcock–Leighton (BL) model, and other dynamo models is not consistently defined. The $\alpha$ effect in the Parker or BL models was developed through a combination of semi-empirical and theoretical approaches to explain observed phenomena. More rigorous theoretical frameworks, such as mean field theory (MFT), the eddy-damped quasi-normal Markovian (EDQNM) approximation, and the direct interaction approximation (DIA), have been employed to calculate these coefficients \citep{1976JFM....77..321P, 1978mfge.book.....M, 1980opp..bookR....K, 2001ApJ...550..824B, Akira2011}. These theories generally suggest that $\alpha$ is related to the residual helicity, $\langle \mathbf{b} \cdot (\nabla \times \mathbf{b}) \rangle - \langle \mathbf{u} \cdot (\nabla \times \mathbf{u}) \rangle$ and the $\beta$ coefficient is considered to be associated with the turbulent energies, $\langle \mathbf{u}^2 \rangle$(MFT, DIA, EDQNM) or $\langle \mathbf{b}^2 \rangle$(DIA, EDQNM, refer to Eqs.~(9.159), (9.183) in \cite{Akira2011} and Eq.~(3.7) in \cite{1976JFM....77..321P} for details).\\

These dynamo models propose that large-scale magnetic fields are amplified by the $\alpha$-effect and suppressed by magnetic $\beta$-diffusion. However, the conclusions regarding the $\alpha$ and $\beta$ effects have been primarily based on leading-order approximations, which reflect only partial aspects of the full dynamics. More rigorous theoretical studies have investigated the possibility of partially negative magnetic diffusivity (\cite{1999GApFD..91..131L}, and references therein), particularly in systems with strong helicity where $\alpha$–$\alpha$ correlations dominate. Also, \cite{2025ApJ...985...18R} proposed a turbulent magnetic diffusion coefficient $\eta_t$ using path integrals. By employing a projection operator, Taylor expansion, second-order moments of velocity, Fourier transforms, and tensor identities, they derived $\eta_t$, which can take negative values, and compared it with simulation results. They found a qualitative result that the magnetic diffusivity decreases when kinetic helicity is present. From a numerical perspective, the Test-Field Method (TFM) was developed to extract the $\alpha$ and $\beta$ coefficients from simulation data \citep{2005AN....326..245S}. Negative magnetic diffusivity has also been observed in liquid sodium experiments \citep{2014PhRvL.113r4501C, 2014NJPh...16g3034G}, where small-scale turbulent fluctuations were found to contribute to this effect in the interior region.\\


Practically, there have been ongoing efforts to incorporate the $\alpha$ and $\beta$ coefficients into solar magnetic field modeling. These efforts mainly aim to reproduce the observed spatial structure and periodicity of the solar magnetic field using $\alpha$ and $\beta$ coefficients extracted from direct numerical simulations (DNS). \citet{2013ApJ...768...16S} obtained the $\alpha_{ij}$ tensor based on the numerical technique introduced by \citet{2011ApJ...735...46R} and the theoretical framework from \citet{1978mfge.book.....M}, using the relation $\boldsymbol{\xi}i = \langle \mathbf{u} \times \mathbf{b} \rangle_i = \alpha_{ij} \overline{B}_j + \beta_{ijk} \partial_k \overline{B}_j$. However, since \citet{2011ApJ...735...46R} neglected the $\beta$ tensor, \citet{2013ApJ...768...16S} employed a simplified form of $\beta$ to represent turbulent diffusivity. They then adopted representative parameter values such as $\alpha_0$, $\eta_0$, and $\Omega_0$. Their approach can be compared with that of \citet{2008A&A...483..949J}, who used a small trial-and-error value of $\alpha$ to ensure numerical stability while reproducing the large-scale features of the solar magnetic field.\\


In our study, we do not aim to reproduce the solar magnetic pattern. Instead, we propose a theoretical and numerical derivation of the $\alpha$ and $\beta$ effects that enables the most accurate reproduction of the DNS results possible. We focus on the electrodynamic aspect of the electromotive force (EMF), $\langle \mathbf{u} \times \mathbf{b} \rangle$, as the coupling coefficient of $\mathbf{B}{_\text{tor}}$ and $\mathbf{B}{_\text{pol}}$. In our prior research \citep{2023ApJ...944....2P, 2025PhRvD.111b3021P}, we successfully derived $\alpha_{\text{EM-HM}}$ and $\beta_{\text{EM-HM}}$ from large-scale magnetic energy $\overline{E}_M (= \langle \overline{B}^2 \rangle / 2)$, and magnetic helicity $\overline{H}_M (= \langle \overline{\mathbf{A}} \cdot \overline{\mathbf{B}} \rangle, ~ \overline{\mathbf{B}} = \nabla \times \overline{\mathbf{A}})$\footnote{\textbf{A} and \textbf{u}: polar vector, \textbf{B} and ${\bm \omega} (=\nabla\times \mathbf{u})$: axial vector $\rightarrow\, \langle \overline{\mathbf{A}} \cdot \overline{\mathbf{B}} \rangle$ and $\langle \mathbf{u} \cdot \bm{\omega} \rangle$: pseudoscalar}, without relying on any artificial assumptions or additional constraints on the system. By utilizing raw simulation data, we generated their profiles and reproduced the evolving large-scale magnetic field, aligning with DNS. Upon confirming that our model's outcomes were consistent with the DNS, we formulated alternative $\beta_{\text{vv-vw}}$, which necessitates turbulent kinetic energy $\langle u^2 \rangle / 2$ and helicity $\langle \mathbf{u} \cdot \nabla \times \mathbf{u} \rangle$. This new expression comprises two components: the conventional $\beta$ effect and an additional term attributable to turbulent kinetic helicity. $\beta_{\text{vv-vw}}$ is to clarify what actually determines diffusion in a magnetized plasma system.\\


In this work, we generated profiles for $\alpha_{\mathrm{EM-HM}}$ and $\beta_{\mathrm{EM-HM}}$, as well as for $\beta_{\mathrm{vv-vw}}$, and used these distinct datasets to reconstruct the large-scale magnetic field $\overline{\mathbf{B}}$. The pair $(\alpha_{\mathrm{EM-HM}}, \beta_{\mathrm{EM-HM}})$ successfully reproduces $\overline{\mathbf{B}}$ throughout the entire evolution. In contrast, $(\alpha_{\mathrm{EM-HM}}, \beta_{\mathrm{vv-vw}})$ reconstructs a reliable $\overline{\mathbf{B}}$ only in the kinematic regime, whereas the field diverges in the nonlinear regime. To address this, we derived a more complete form of $\beta$ that includes turbulent magnetic effects, namely $\beta_{\mathrm{vv-vw}} + \beta_{\mathrm{bb+jb}}$, and used it to reconstruct $\overline{\mathbf{B}}$. The inclusion of turbulent magnetic effects clearly suppresses the divergence of $\overline{\mathbf{B}}$. Then, we investigated how the $\alpha$ and $\beta$ effects influence the amplification of $\overline{\mathbf{B}}$. By varying their weights from $-100$\% to $+100$\%, we analyzed the sensitivity of $\overline{\mathbf{B}}$ and compared the results with DNS data. The comparison reveals that the $\beta$ effect plays a dominant role in amplifying the magnetic field, while the $\alpha$ effect primarily shapes the profile of the saturated field.\\

This article is structured into five chapters. Chapter 2 briefly discusses the evolution of the magnetic field in a rotating spherical plasma system like the Sun. Unlike Parker’s or Babcock–Leighton (BL) models, our approach is more directly based on electromagnetic interactions described by the magnetic induction equation. However, constructing a detailed solar model is not the main goal of this work. Our primary objective is to identify the $\alpha$ and $\beta$ effects that account for the solar magnetic field. In the third chapter, we discuss the basic MHD dynamo theory. The detailed derivations are included in the Appendix to improve consistency and readability. The fourth chapter presents the simulation approach and numerical results, which are used to verify the theoretical findings. We reproduce the large-scale magnetic field by adjusting the weights of the $\alpha$ and $\beta$ effects. Finally, the article concludes with a summary.\\

\begin{figure*}
\centering
    {
   \subfigure[In the Southern Hemisphere, \textbf{positive kinetic helicity} $ \langle \mathbf{u} \cdot \boldsymbol{\omega} \rangle $ is generated by the combined action of the Coriolis force and buoyancy.]{
     \includegraphics[width=15 cm]{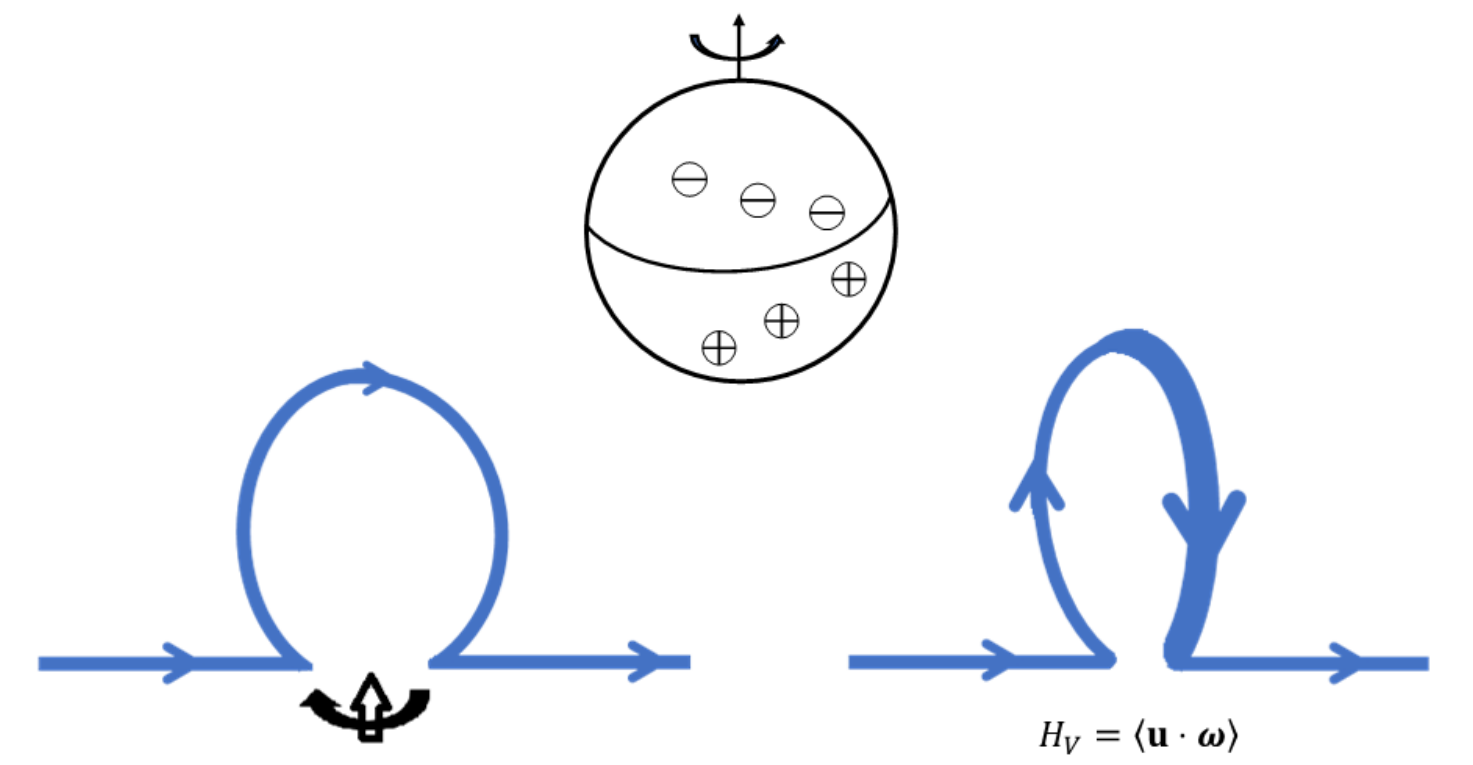}
     \label{f2a}
    }\hspace{-13 mm}
   \subfigure[$\langle \overline{\mathbf{j}}_0\cdot \overline{\mathbf{B}}_0\rangle<0$ and $\langle \mathbf{j}_1\cdot \mathbf{b}_{ind}\rangle>0$]{
   \includegraphics[width=17 cm]{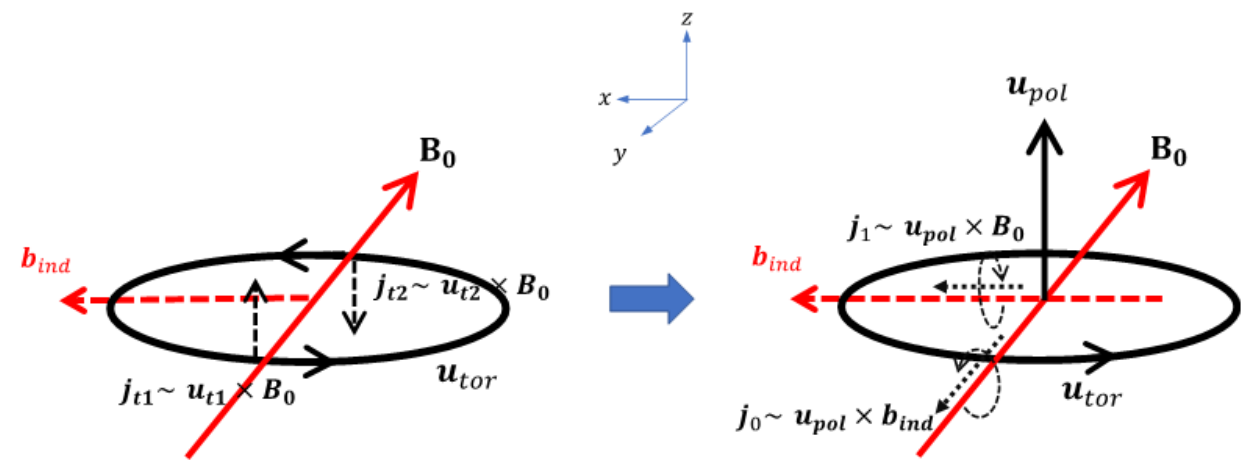}
     \label{f2b}
   }
}
\caption{(a) Right handed $\langle \mathbf{u}\cdot \mathbf{\omega}\rangle$ in southern hemisphere. The plasma flux, buoyed toward the surface, is twisted clockwise by the Coriolis force, generating right-handed kinetic helicity as a result.  Note that kinetic effects generate kinetic helicity, not the magnetic flux. (b) Left handed $\langle \mathbf{j}_0\cdot \mathbf{B}_0\rangle$ and right handed $\langle \mathbf{j}_1\cdot \mathbf{b}_{ind}\rangle$ by right handed $\langle \mathbf{u}\cdot \mathbf{\omega}\rangle$.}
\end{figure*}

\section{Overview of magnetic evolution in a Rotating spherical system}
In this section, we qualitatively explain, based on electromotive theory, the process in which $\mathbf{B}_{tor}$ is converted to $\mathbf{B}_{pol}$ by current density in a rotating body, and $\mathbf{B}_{pol}$ is converted back to $\mathbf{B}_{tor}$ by differential rotation.

\subsection{Temporal Evolution of Solar Magnetic Field}

In Figure 1(a), we assume that $B_{\text{pol}}$, flowing from north to south, is distorted by the sphere's differential rotation. This rotation then leads to the formation of ($B_{\text{tor}}$), which becomes most concentrated near the equator. Due to the Coriolis force and buoyancy in the rotating celestial body, left-handed (negative) kinetic helicity ($\langle \mathbf{u}\cdot \omega\rangle,\,\omega=\nabla\times {\bf u})$ is generated in the northern hemisphere, while right-handed (positive) kinetic helicity forms in the southern hemisphere (refer to Fig.~\ref{f2a}). In the numerical code, a series of these turbulent kinetic motions is included in the forcing term $\mathbf{f}(\mathbf{r},\,t)$ in the momentum equation. Other convective motions are represented by the large-scale velocity field $\overline{U}$.\\

As Figure 1(b) shows, in the northern hemisphere, $\mathbf{J}$ induced by left-handed kinetic helicity is in the same direction as the magnetic flux. On the contrary, $\mathbf{J}$ is induced in the opposite direction in the southern hemisphere. Due to the opposite direction of $\mathbf{B}_{\text{tor}}$ in both hemispheres, the current densities in the northern and southern hemispheres flow from west to east. We will discuss the generation of $\mathbf{J}$ aligned with $\mathbf{B}_{tor}$ due to kinetic helicity in Fig.~\ref{f2b}.\\

Figure~1(c) illustrates that these current densities, in turn, induce new magnetic fields $ \mathbf{B}_{\text{pol}0} $ encircling $ \mathbf{B}_{\text{tor}} $, as described by Maxwell theory ($ \nabla \times \mathbf{B} = \mu_0\mathbf{J} $), not by Parker's model. As a result, $\mathbf{B}_{\text{tor}}$ and $\mathbf{B}_{\text{pol}0}$ generate right-handed magnetic helicity in the northern hemisphere and left-handed magnetic helicity in the southern hemisphere(Figs.~\ref{f4a}, \ref{f4b}, $\alpha$ effect). The fields $\mathbf{B}_{\text{pol}0}$ reconnect to form the poloidal field ($\mathbf{B}_{\text{pol}}$).\\

In Figure 1(d), these small poloidal fields combine to form a large-scale $\mathbf{B}_{\text{pol}}$ that flows from the northern pole to the southern pole ($N\rightarrow S$). However, the large-scale $\mathbf{B}_{\text{pol}}$ formed beneath $\mathbf{B}_{\text{tor}}$ flows in the opposite direction, i.e., from south to north ($S\rightarrow N$). This $\mathbf{B}_{\text{pol}}$ forms a new toroidal field ($\mathbf{B}_{\text{tor}2}$) through differential rotation($\alpha-\Omega$ effect). $\mathbf{B}_{\text{tor}2}$ flows in the opposite direction to the original $\mathbf{B}_{\text{tor}}$, as depicted in Figure 1(e)(dashed line). Meanwhile, the strongest $\mathbf{B}_{\text{tor}}$ near the solar surface has fewer or virtually no factors for further amplification compared to $\mathbf{B}_{\text{tor}2}$ which is still strengthened by differential rotation. Note that the polarity of $\langle \mathbf{J}_{ind}\cdot \mathbf{B}_{tor2}\rangle$ is opposite to that of $\langle \mathbf{J}_{ind}\cdot \mathbf{B}_{tor}\rangle$.\\

As $\mathbf{B}_{\text{tor}2}$ strengthens, the plasma density within the magnetic flux tube decreases, making it lighter and causing it to rise to the surface. There, it reconnects with the existing magnetic flux $\mathbf{B}_{\text{tor}}$, canceling out the magnetic fields. The amplified, oppositely directed magnetic flux beneath the surface continues to rise, eventually reaching the surface and exhibiting reversed polarity, as shown in Figs.~1(f) and 1(g). In the new cycle depicted in Figure 1(h), the current densities in both hemispheres flow from east to west, driven by the polarities of kinetic helicity. This process repeats every 11 years, accounting for the polarity reversal of the solar surface magnetic flux and the brief disappearance of magnetic fields between them. Nonetheless, $\langle \mathbf{J}_{\mathrm{ind}} \cdot \mathbf{B}_{\mathrm{tor}} \rangle$ in the southern (northern) hemisphere is always left (right) handed, regardless of the reversal of the magnetic field.\\

\subsection{The Role of Kinetic Helicity in Magnetic Field Amplification via $\alpha$ and $\beta$ Effects}

Fig.~\ref{f2a} shows the kinetic helicity generated in a rotating plasma sphere. In the solar convection zone ($0.7R_{\bigodot}-1.0R_{\bigodot}$), numerous tube-shaped plasma flux structures exist. The plasma inside the tube is pushed outward due to the balance between magnetic pressure and thermal pressure, making it relatively lighter than the surrounding medium and causing it to rise toward the solar surface. During this process, a portion of the plasma eddy stretches, reducing its density further and increasing its buoyancy, which leads to the deformation of the tube structure. At this stage, the Coriolis force causes the circular loop to rotate in a clockwise direction, generating right-handed kinetic helicity in the Southern Hemisphere. Conversely, in the Northern Hemisphere, left-handed kinetic helicity is generated. This kinetic helicity interacts with the magnetic field, inducing new magnetic fields (right handed polarity), as illustrated in Fig.~\ref{f2b}.\\

In Fig.~\ref{f2b}, the left panel illustrates a circular structure of a plasma turbulence eddy through which a magnetic field, $\mathbf{B}_0$, permeates, influencing both the eddy’s structure and energy distribution. Plasma motions, labeled $\mathbf{u}_{t1}$ and $\mathbf{u}_{t2}$, interact with $\mathbf{B}_0$ to produce current densities, $\mathbf{j}_{t1} (\hat{z})$ and ${\bf j}_{t2} (-\hat{z})$. According to Ampère's law, these current densities induce a magnetic field, $\mathbf{b}_{\text{ind}} (\hat{x})$, creating magnetic diffusion through the relationship $\nabla \times \mathbf{j}_{t} \sim \nabla \times (\nabla \times \mathbf{b}) \sim -\nabla^2 \mathbf{b}$. This process of magnetic field induction occurs continuously, ultimately leading to the weakening of the original magnetic field. These sequential processes explain the magnetic diffusion $\beta$ effect due to plasma turbulence fluctuations $\beta\sim \int \langle u^2 \rangle d\tau$. This is the turbulent magnetic diffusion in small scale dynamo.\\

However, if there is buoyancy, an additional poloidal velocity $\mathbf{u}_{\text{pol}}$ appears. This component interacts with $\mathbf{b}_{\text{ind}}$ to generate a current density $\mathbf{u}_{pol}\times \mathbf{b}_{ind}\sim\mathbf{j}_{0} (\hat{y})$ along $\mathbf{B}_0$, which in turn induces a toroidal magnetic field $\mathbf{b}_{\text{tor}}$(dotted circle). $\mathbf{B}_{0}$ (acting as a poloidal field) and $\mathbf{b}_{\text{tor}}$ form left-handed magnetic helicity, $H_{M1}$($\alpha$ effect). Simultaneously, $\mathbf{u}_{\text{pol}}$ can interact with $\mathbf{B}_{0}$ to produce another current density, $\mathbf{j}_{1}$, with $\mathbf{b}_{\text{ind}}$ and $\mathbf{j}_{1}$ forming right-handed magnetic helicity, $H_{M2}$. This sequence of processes indicates that the $\alpha$ process presumes the preceding magnetic diffusion $ \beta $.\\

In this figure, $\mathbf{B}_0$, $\mathbf{u}_{\text{pol}}$, and $\mathbf{b}_{\text{ind}}$ are depicted as intersecting at the same point. However, while $\mathbf{b}_{\text{ind}}$ and $\mathbf{u}_{\text{pol}}$ intersect each other, $\mathbf{B}_0$ and $\mathbf{u}_{\text{pol}}$ can be separated. Statistically, $\mathbf{j}_0$ becomes greater than $\mathbf{j}_1$. This left handed $H_{M1}$ can be interpreted as the magnetic helicity generated by right handed kinetic helicity and cascaded inversely to larger scales. In contrast, right handed $H_{M2}$ can be understood as the magnetic helicity produced to conserve the total magnetic helicity within the system. This reasoning aligns with the theoretical prediction that kinetic helicity generates magnetic helicity of opposite polarity, which is then amplified and transferred to larger scales, while magnetic helicity of the same polarity is generated in the small-scale regime.\\

Going back to Fig.~\ref{f1}, as a thought experiment, consider the case where the current density generated in Fig.~1(b) is reversed in direction. This represents a possible case of a magnetically driven system. Then, the corresponding poloidal magnetic field $ {\bf B}_{\text{pol}} $ beneath $ {\bf B}_{\text{tor}} $ would also reverse its direction (Fig.~1(d), from south-to-north to north-to-south). In this case, differential rotation would continue to generate toroidal fields $ {\bf B}_{\text{tor2}} $ of the same direction (west to east) beneath the original one, leading to continuous amplification of ${\bf B}_{tor}$. Eventually, this unconstrained amplification would lead to a magnetic catastrophe. The periodic reversal of the magnetic field is indispensable for the stable evolution of a rotating plasma system.\\

The model in Fig.~\ref{f1} is based on the helical kinetic-forcing dynamo. Therefore, the polarities of the kinetic helicity and magnetic helicity are opposite to each other. Furthermore, the magnetic helicities at large and small scales also exhibit opposite polarities. However, if the magnetic flux tube is primarily twisted by buoyancy and the Coriolis force—i.e., a magnetically forced dynamo—then left-handed (right-handed) magnetic helicity is generated in both the large-scale and small-scale ranges in the northern (southern) hemisphere. There is no polarity reversal within the same hemisphere.\\

Then, we may wonder what the magnetic helicities in the Sun actually are. \cite{Pipin_2014} showed that the global (large-scale) magnetic helicity is positive in the northern hemisphere and negative in the southern hemisphere. They used two different data sets: one from SOHO and the other from SOLIS. They also emphasized that the opposite signs—i.e., negative (positive) magnetic helicity in the northern (southern) hemisphere—are typically associated with active regions, such as sunspots\citep{2010MNRAS.402L..30Z}. In fact, the global-scale magnetic field is comparable in size to the solar hemisphere itself. According to their observations, the magnetic helicity polarity in each hemisphere appears to change over time.\\

\section{Numerical Approach}

\subsection{Basic magnetohydrodynamic equations}
The solar dynamo can be described by a set of nonlinear magnetohydrodynamic (MHD) equations that govern the dynamics of the electrically conducting magnetized plasma within the plasma systems.
\begin{eqnarray}
\frac{D \rho}{Dt}&=&-\rho {\bf \nabla} \cdot {\bf U},\label{continuity equation_original}\\
\frac{D {\bf U}}{Dt}&=&-{\bf \nabla} \mathrm{ln}\, \rho + \frac{1}{\rho}(\nabla\times{\bf B})\times {\bf B}+\nu\bigg({\bf \nabla}^2 {\bf U}+\frac{1}{3}{\bf \nabla} {\bf \nabla} \cdot {\bf U}\bigg)+\mathbf{f}(\mathbf{r},\,t)
\label{momentum equation_original}\\
\frac{\partial \mathbf{B}}{\partial t}&=&\nabla \times \langle \mathbf{U}\times \mathbf{B}\rangle +\eta \nabla^2\mathbf{B}
\label{magnetic induction equation_original}
\end{eqnarray}
($\rho$, $U$, $B$, and $D/Dt(=\partial / \partial t + {\bf U} \cdot {\bf \nabla}$) indicate the density, velocity field, magnetic field, and Lagrangian time derivative in order. And, $\nu$ \& $\eta$ are kinematic viscosity and magnetic diffusivity respectively.)\\

These equations encapsulate the interactions between plasma flow, magnetic fields, and thermodynamic processes, providing a framework for understanding the generation and evolution of the density, velocity, and magnetic fields. The dynamo mechanism functions through the combined effects of differential rotation, convective motions, turbulent effects, diffusion, and forcing sources $\mathbf{f}(\mathbf{r},\,t)$, leading to the amplification of magnetic fields. This forcing source refers to internal and external energy sources such as tidal forces, buoyancy, Coriolis force, electromagnetic force, and gravity, and it can be defined differently depending on the context. However, in this paper, it is used specifically to represent the supply of kinetic helicity generated by buoyancy and the Coriolis force. These foundational equations are commonly applied in direct numerical simulations (DNS) or theoretical analyses.\\


Particularly when the fields in the system exhibit helicity, $\langle {\bf F}\cdot \nabla \times \mathbf{F}\rangle =\lambda \langle F^2\rangle$, the magnetic induction equation [Eq.~(\ref{magnetic induction equation_original})] is modified (rederived) to include $\alpha$ and $\beta$ coefficients, alongside the large-scale magnetic field $\overline{\mathbf{B}}$ and plasma velocity $\overline{\mathbf{U}}$.\footnote{$\alpha$ disappears without helicity.}
\begin{eqnarray}
\frac{\partial \overline{\mathbf{B}}}{\partial t}&=&\nabla \times \big(\overline{\mathbf{U}}\times \overline{\mathbf{B}}+\langle \mathbf{u}\times \mathbf{b} \rangle\big)+\eta \nabla^2\overline{\mathbf{B}}\\
&\sim &\nabla \times (\alpha \overline{\mathbf{B}}-\beta\nabla\times \overline{\mathbf{B}})+\eta \nabla^2\overline{\mathbf{B}} \sim\,\alpha\overline{\mathbf{J}}+(\beta+\eta) \nabla^2\overline{\mathbf{B}}
\label{magnetic induction equation_alpha_beta}
\end{eqnarray}
For an isotropic and homogeneous system, $\overline{\mathbf{U}}$ can be removed by applying a Galilean transformation. However, in the presence of shear, the mean flow cannot be transformed out. For a rotating spherical system, it is more convenient to express Eq.~(\ref{magnetic induction equation_alpha_beta}) in curvilinear coordinates \cite{Akira2011, 2014ARA&A..52..251C}:
\begin{eqnarray}
&&\frac{\partial \overline{\mathbf{A}}_{\phi}}{\partial t}+\frac{1}{\sigma}\overline{\mathbf{U}}_p\cdot\nabla(\sigma \overline{\mathbf{A}}_{\phi})=\alpha \overline{\mathbf{B}}_{\phi}+(\beta+\eta)\bigg(\nabla^2-\frac{1}{\sigma^2}\bigg)\,\overline{\mathbf{A}}_{\phi},
\label{magnetic induction equation_poloidal}\\
&&\frac{\partial \overline{\mathbf{B}}_{\phi}}{\partial t}+\sigma\overline{\mathbf{U}}_p\cdot\nabla\bigg(\frac{\overline{\mathbf{B}}_{\phi}}{\sigma}\bigg)=\alpha(\nabla \times \overline{\mathbf{B}}_p)_{\phi}+\sigma(\overline{\mathbf{B}}_p\cdot\nabla)\frac{\overline{\mathbf{U}}_{\phi}}{\sigma}+(\beta+\eta)\,\bigg(\nabla^2-\frac{1}{\sigma^2}\bigg)\,\overline{\mathbf{B}}_{\phi},
\label{magnetic induction equation_toroidal}
\end{eqnarray}
where $\overline{\mathbf{B}}=\overline{B}_{\phi}\hat{e}_{\phi}+\overline{\mathbf{B}}_p$, $\overline{\mathbf{B}}_p=\nabla\times (\overline{A}_{\phi}\hat{e}_{\phi})$, and $\sigma=r\,sin\,\theta$. This equation is not a semi-empirical one based on assumptions, but can be directly derived through a standard coordinate transformation process from cartesian to spherical coordinates \citep{2005mmp..book.....A}.\\

The large-scale magnetic field $\overline{\mathbf{B}}$ (or $\overline{\mathbf{A}}$) cannot be sustained without the presence of $\alpha$ or $\beta$, both of which implicitly incorporate the effects of the external forcing source $\bf f$. The poloidal and toroidal components of the magnetic field are coupled through the $\alpha$ effect, while both components are influenced by $\beta$-induced diffusion. The gradient $\nabla \overline{U}_{\phi}$ arises from the rotation of the spherical system, whereas the $\alpha$ and $\beta$ coefficients originate from turbulent (helical) plasma motions and magnetic fields. Since the sign of helicity is opposite in the Northern and Southern Hemispheres, it is essential to examine how the $\alpha$ and $\beta$ effects vary across hemispheres and to verify whether the resulting magnetic field amplification aligns with theoretical predictions.\\


\subsection{Numerical Method}
We used the \texttt{PENCIL CODE} to perform our numerical simulations. This code solves Eqs.~(\ref{continuity equation_original})-(\ref{magnetic induction equation_original}) within a periodic cube of size $ (2\pi)^3 $, discretized into a grid of $ 400^3 $ points. The velocity and magnetic fields are scaled by the sound speed, $ c_s $, and $ (\rho_0\,\mu_0)^{1/2}c_s $, respectively. These scalings follow from the relations $ E_M \sim B^2/\mu_0 $ and $ E_V \sim \rho_0 U^2 $, where $ \mu_0 $ and $ \rho_0 $ represent the magnetic permeability of free space and the initial density. It’s worth noting that the plasma system is weakly compressible, meaning $ \rho \sim \rho_0 $. The system is forced with $ {\bf f}(\mathbf{r},\,t)=N\,{\bf f}(t)\,\exp\left[i\,{\bf k}_f(t)\cdot {\bf x} + i\phi(t)\right]$, which is attached to the momentum equation Eq.~(\ref{momentum equation_original}). $ N $ is a normalization factor, $ {\bf f}(t) $ is the forcing magnitude, and $ {\bf k}_f(t) $ represents the forcing wave number. At each time step, the code randomly selects one of 20 vectors from the $ {\bf k}_f $ set. For simplicity, we set $ c_s $, $ \mu_0 $, and $ \rho_0 $ to 1, making the equations dimensionless.\\

The forcing function ${\bf f}(t)$ is defined as $f_0\mathbf{f}_k(t)$:
\begin{eqnarray}
{\bf f}_k(t)=\frac{i\mathbf{k}(t)\times (\mathbf{k}(t)\times \mathbf{\hat{e}})-\lambda |{\bf k}(t)|(\mathbf{k}(t)\times \mathbf{\hat{e}})}{k(t)^2\sqrt{1+\lambda^2}\sqrt{1-(\mathbf{k}(t)\cdot \mathbf{e})^2/k(t)^2}}.
\label{forcing amplitude fk}
\end{eqnarray}
The wavenumber `$k$' is defined as $ 2\pi/l $. A value of $ k = 1\,(l=2\pi)$ corresponds to the large-scale regime, while $ k > 2 $ refers to the wavenumber in the small (turbulent) scale regime. The parameter $ \lambda = \pm 1 $ generates a fully right- ($ \lambda = +1 $) or left-handed ($ \lambda = -1 $) helical field, described by $\nabla \times {\bf f}_k \rightarrow i \mathbf{k} \times \mathbf{f}_k \rightarrow \pm k \mathbf{f}_k$. The choice of $ \lambda = +1 $ represents right-handed polarization, corresponding to the southern hemisphere, while $ \lambda = -1 $ represents left-handed polarization for the northern hemisphere. Here, $ \mathbf{\hat{e}} $ denotes an arbitrary unit vector. We applied fully helical kinetic energy ($ \lambda = \pm 1 $) at $ \langle k \rangle_{\mathrm{ave}} \equiv k_f \sim 5 $. And, we used $ f_0 = 0.07 $ and $ \nu = \eta = 0.006 $. Note that Reynolds' rule is not applied to this energy source: $ \langle f \rangle \neq 0 $. Notably, an initial seed magnetic field of $B_0 \sim 10^{-4}$ was introduced into the system. However, the influence of this seed field diminishes rapidly due to the presence of the forcing function and the lack of memory in the turbulent flow.

\subsection{Numerical Results}
\begin{figure*}
    {
   \subfigure[$\int E_V(k) dk =  \langle u^2\rangle/2$]{
     \includegraphics[width=9.0 cm]{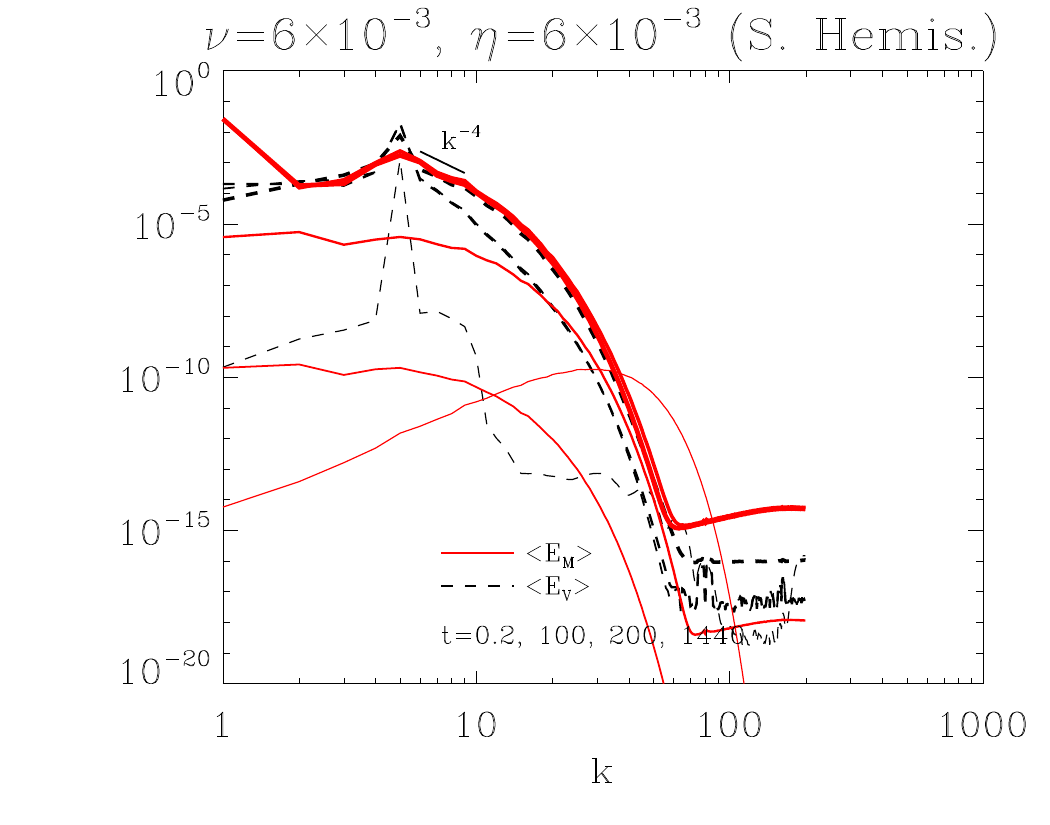}
     \label{f3a}
    }\hspace{-13 mm}
   \subfigure[$\int E_M(k) dk =  \langle b^2\rangle/2$]{
   \includegraphics[width=9.0 cm]{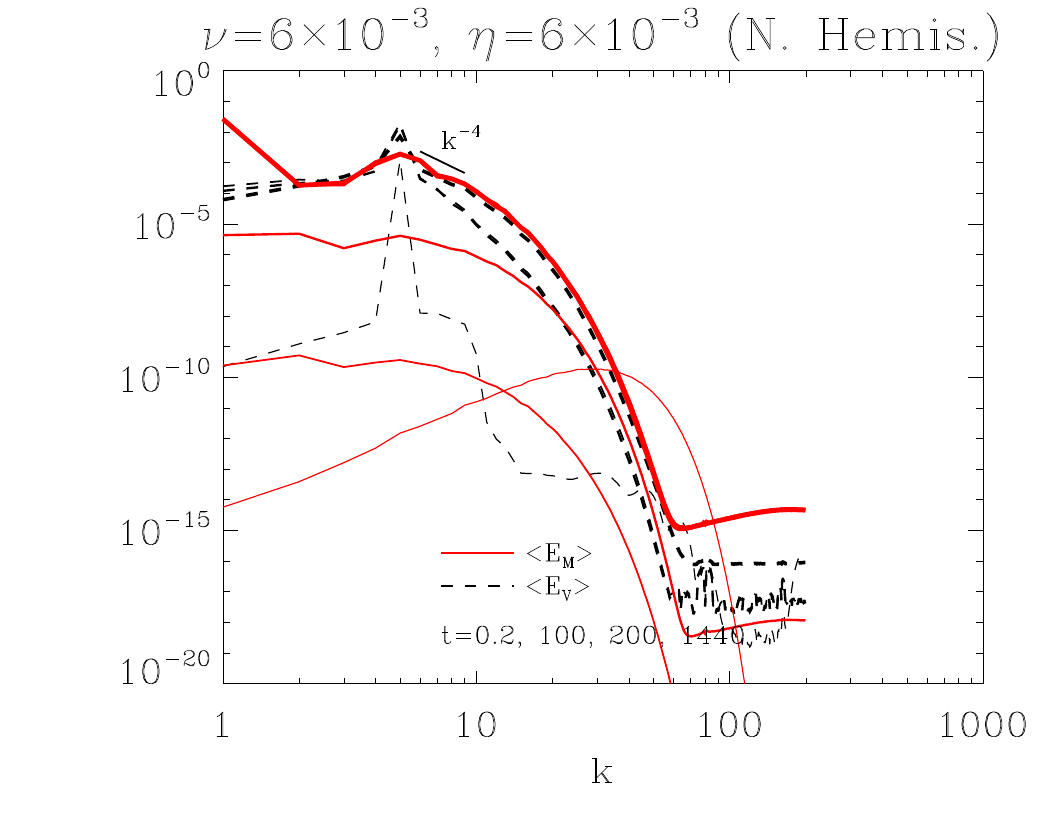}
     \label{f3b}
   }\hspace{-13 mm}
   \subfigure[]{
   \includegraphics[width=9.0 cm]{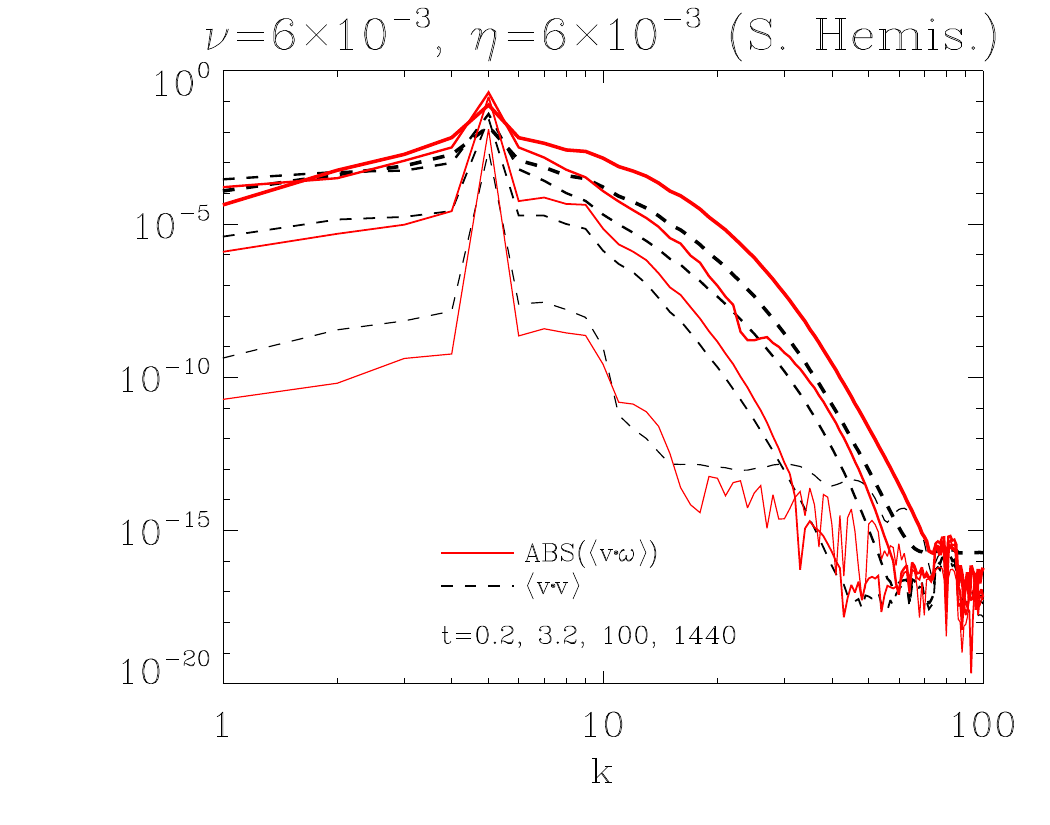}
     \label{f3c}
   }\hspace{-13 mm}
   \subfigure[]{
     \includegraphics[width=9.0 cm]{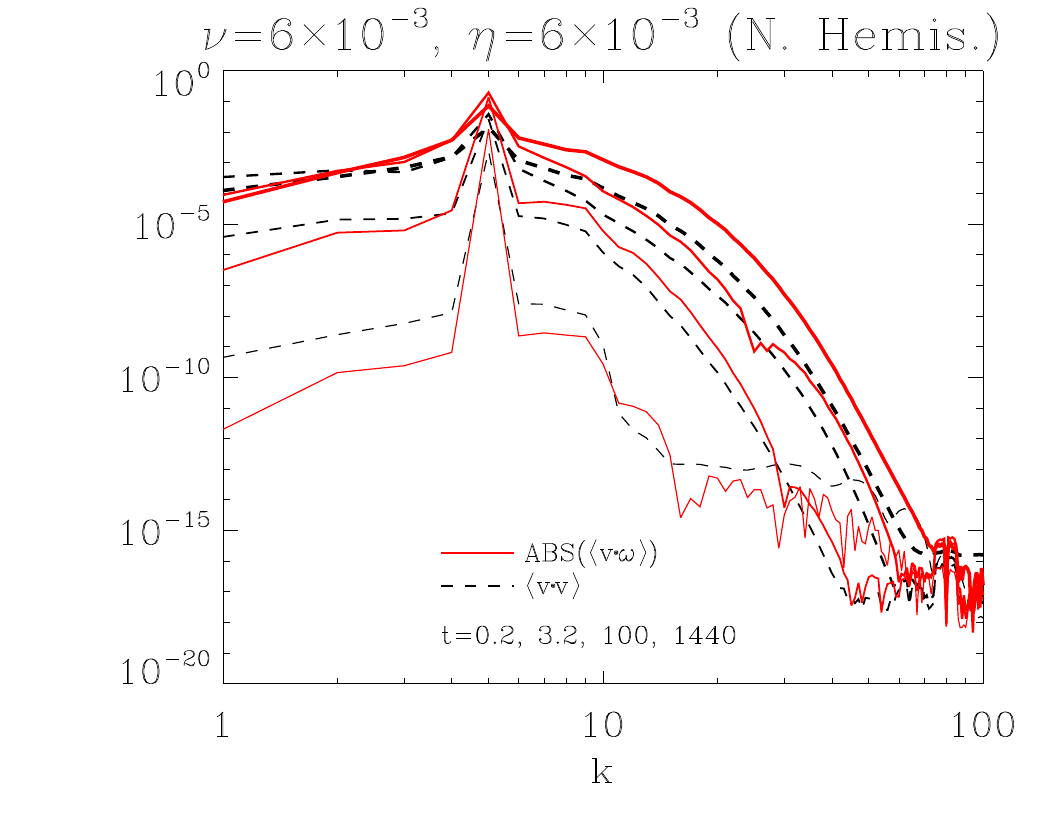}
     \label{f3d}
   }\hspace{-13 mm}
      \subfigure[]{
   \includegraphics[width=9.2 cm]{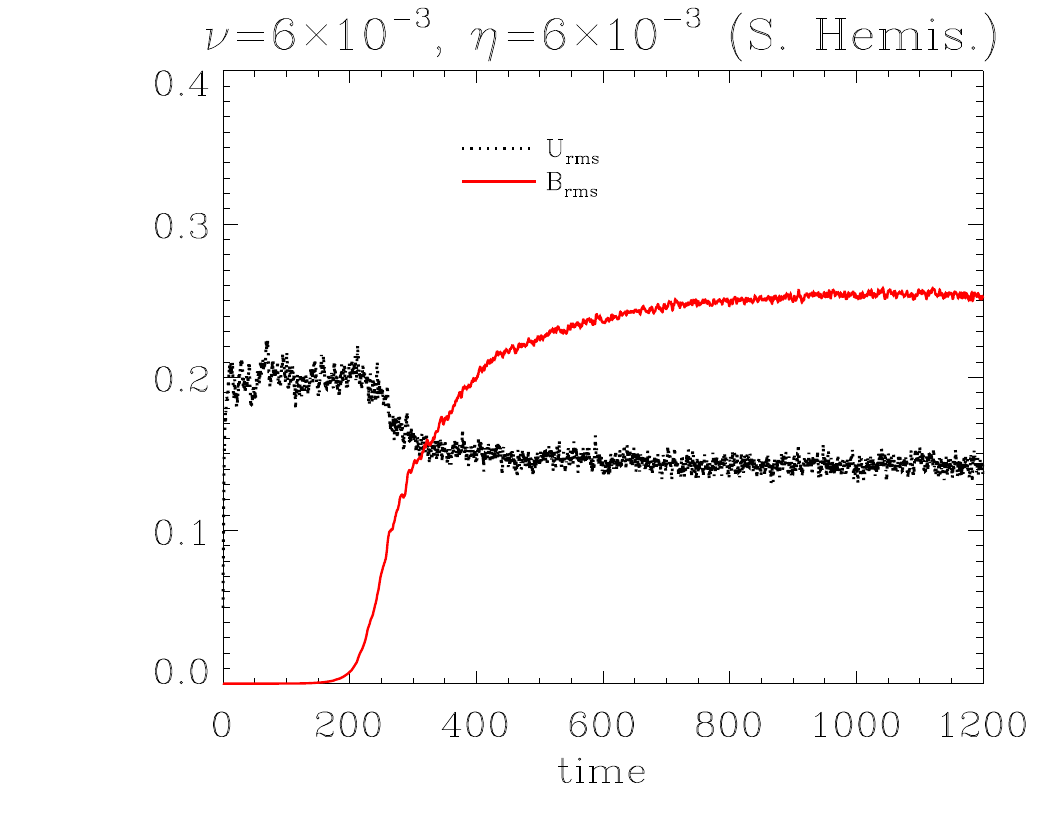}
     \label{f3e}
   }\hspace{-8 mm}
   \subfigure[]{
     \includegraphics[width=9.2 cm]{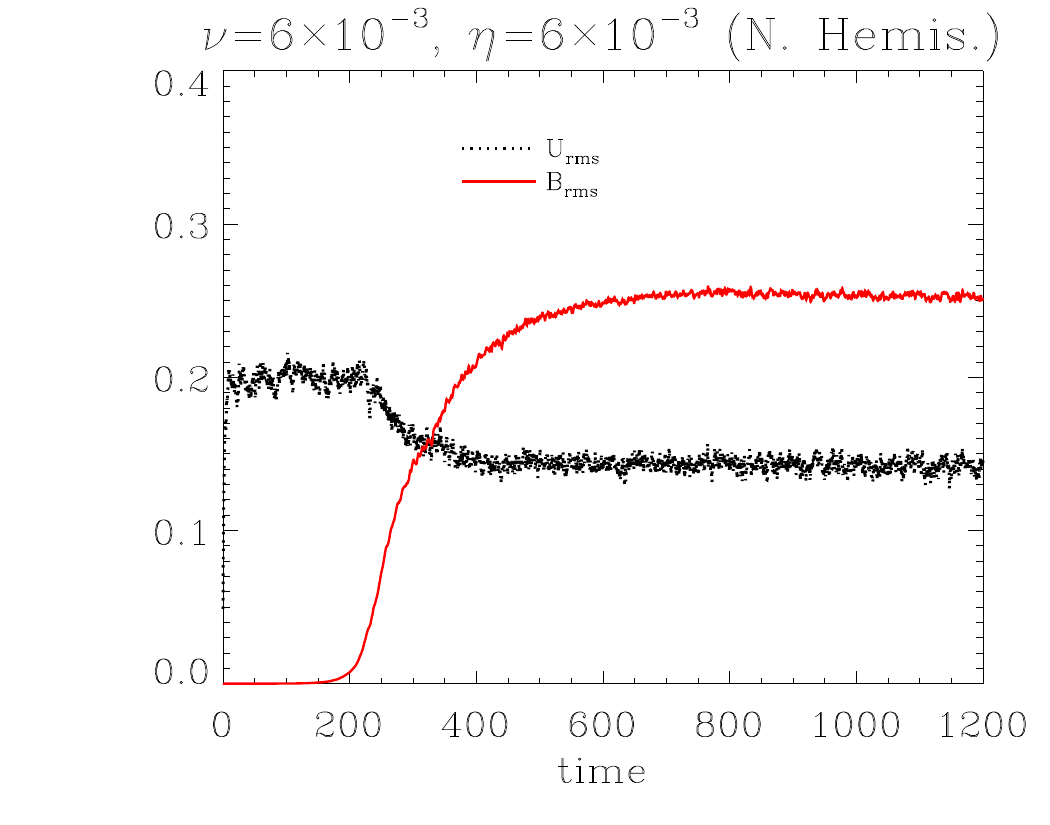}
     \label{f3f}
   }
}
\caption{Left panel: southern hemisphere (positive kinetic helicity), right panel: northern hemisphere (negative kinetic helicity).}
\end{figure*}

Figure 3 shows the energy, helicity, and averaged velocity and magnetic field in a kinetically driven plasma system. The left panel presents the southern hemisphere, which generates positive (right-handed) kinetic helicity, while the right panel shows the northern hemisphere, which generates negative (left-handed) kinetic helicity. Figs.~\ref{f3a} and \ref{f3b} depict the magnetic energy spectrum $E_M$ (red solid line) and kinetic energy spectrum $E_V$ (black dashed line) in Fourier space at times $ t = 0.2, 100, 200, $ and $ 1440 $, showing the evolution of the energy spectra. As observed, the energy spectra for the southern and northern hemispheres match precisely. Both panels illustrate that the kinetic energy $ E_V $ at the forcing scale $ k=5 $ is converted into magnetic energy $ E_M $, which is subsequently inverse cascaded to the large scale $ k=1 $. At this scale, $ E_V $ is significantly low, indicating an insufficient inverse cascade of kinetic energy. Additionally, each energy spectrum $k>5$ decreases more steeply than the Kolmogorov scale $ k^{-5/3} $, implying that the kinetic energy $ E_V $ from the forcing scale does not fully cascade down to smaller plasma eddy regions, a limitation arising from the low magnetic Reynolds number $ Re_M = 261 $.\\

In Figs.~\ref{f3c} and \ref{f3d}, the kinetic helicity $H_V(=\langle \mathbf{u}\cdot \nabla \times \mathbf{u} \rangle)$ and kinetic energy $2E_V(=\langle u^2 \rangle)$ spectra over time are illustrated, with kinetic helicity represented by the red solid line and kinetic energy by the black dashed line. While both Hemispheres exhibit similar trends, there is a clear difference in the sign of kinetic helicity. In Fig.~\ref{f3c}, which represents the southern hemisphere, the kinetic helicity is positive, whereas in Fig.~\ref{f3d}, corresponding to the northern hemisphere, it is negative. Therefore, their absolute values are used for their comparison. Moreover, there is a notable difference between the kinetic energy and kinetic helicity spectra, particularly in their direction of migration. In both hemispheres, for wave numbers $k > 2$ (small-scale regime), the kinetic helicity exceeds the kinetic energy. However, as $k \rightarrow 1$ (large-scale regime), the kinetic energy becomes dominant. Such an unbalanced distribution of $H_V$ and $E_V$ is closely related to the $\alpha$ and $\beta$ effects. An excess of $H_V$ and a deficiency of $E_V$ in small-scale regimes are more favorable for enhancing these $\alpha$ and $\beta$ effects.\\

Figs.~\ref{f3e} and \ref{f3f} illustrate the average velocity and magnetic field in both the Northern and Southern Hemispheres. Using the data from Figs.~\ref{f3a} and \ref{f3b}, we calculated $ U_{\text{rms}} = \sqrt{\sum_k 2E_V} $ and $ B_{\text{rms}} = \sqrt{\sum_k 2E_M} $. The resulting values are $ U_{\text{rms}} \approx 0.14 $ and $ B_{\text{rms}} \approx 0.25 $, which correspond to $ Re \approx 146 $ and $Re_M \approx 241$. $Re_M > Re $ indicates that a greater portion of energy is transferred to smaller-scale magnetic eddy regions, which aligns with typical outcomes of the dynamo process. However, even though $ E_M $ and $ E_V $ are not in equilibrium, particularly when the magnetic Prandtl number $ Pr_M \gg1$, the dissipation wavenumbers, $ k_{\text{max},\,V}$ and $ k_{\text{max},\,M} $, remain the same. In the absence of $ E_V $, the transport of $E_M $ is not possible.\\



\begin{figure*}
    {
   \subfigure[$\langle \mathbf{u}\cdot \bm{\omega}\rangle>0$]{
     \includegraphics[width=9.2 cm]{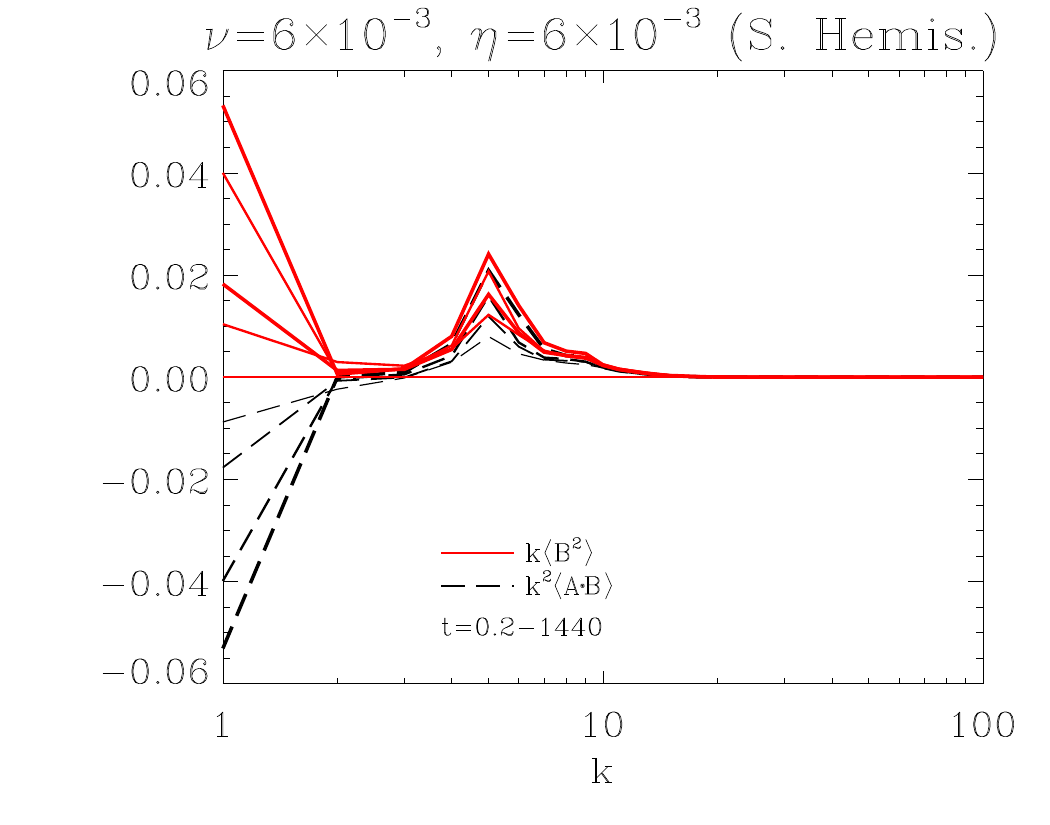}
     \label{f4a}
    }\hspace{-13 mm}
   \subfigure[$\langle \mathbf{u}\cdot \bm{\omega}\rangle<0$]{
   \includegraphics[width=9.2 cm]{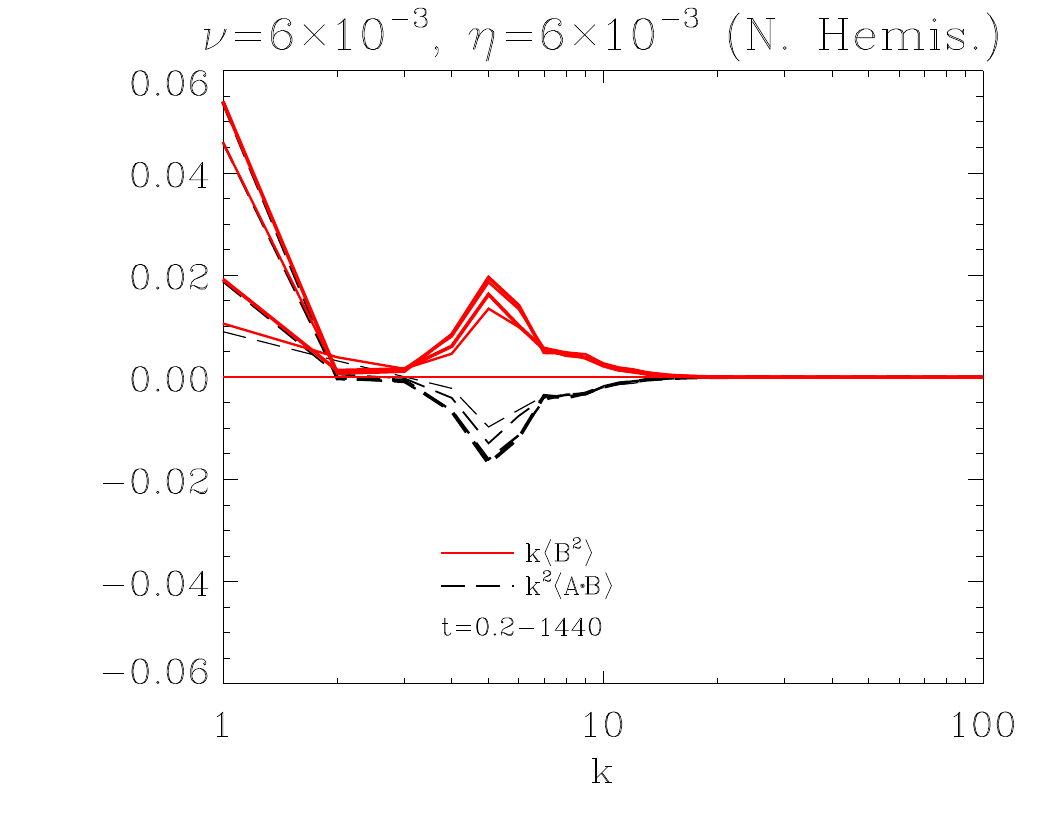}
     \label{f4b}
   }\hspace{-13 mm}
   \subfigure[$f_{hk}=\langle \mathbf{u}\cdot \mathbf{\omega}\rangle/k \langle u^2\rangle$]{
   \includegraphics[width=9.2 cm]{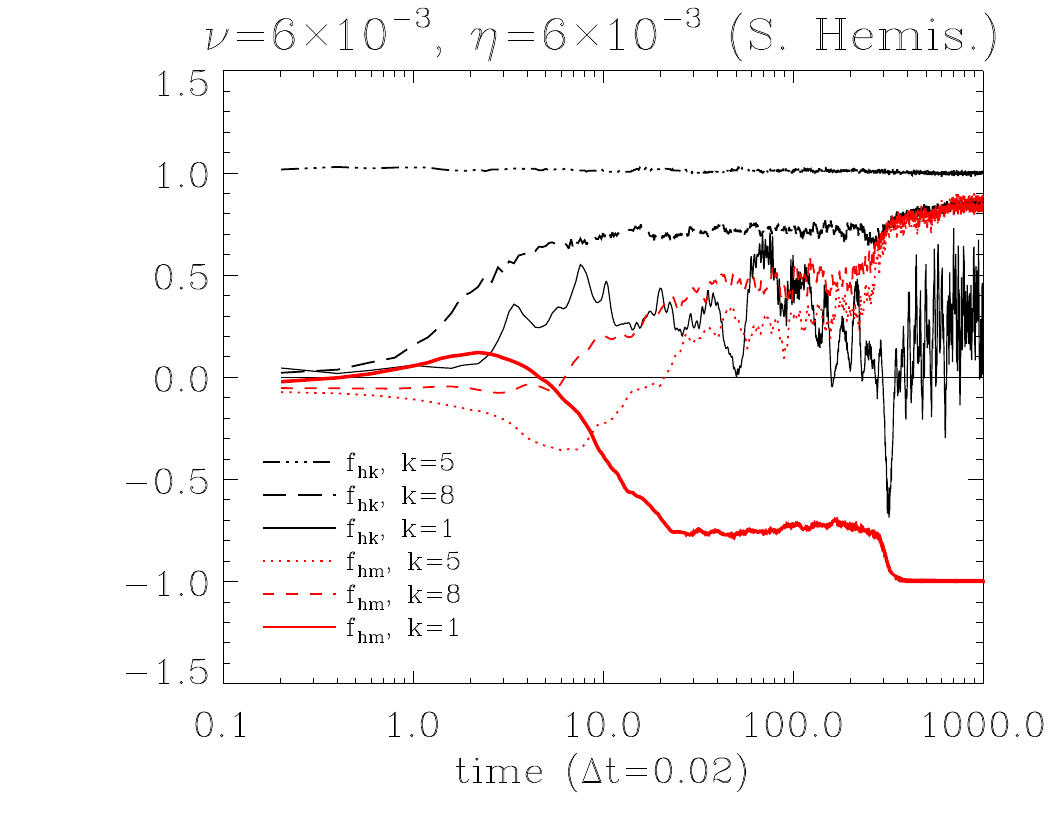}
     \label{f4c}
   }\hspace{-10 mm}
   \subfigure[$f_{hm}=k\langle \mathbf{a}\cdot \mathbf{b}\rangle/ \langle b^2\rangle$]{
   \includegraphics[width=9.2 cm]{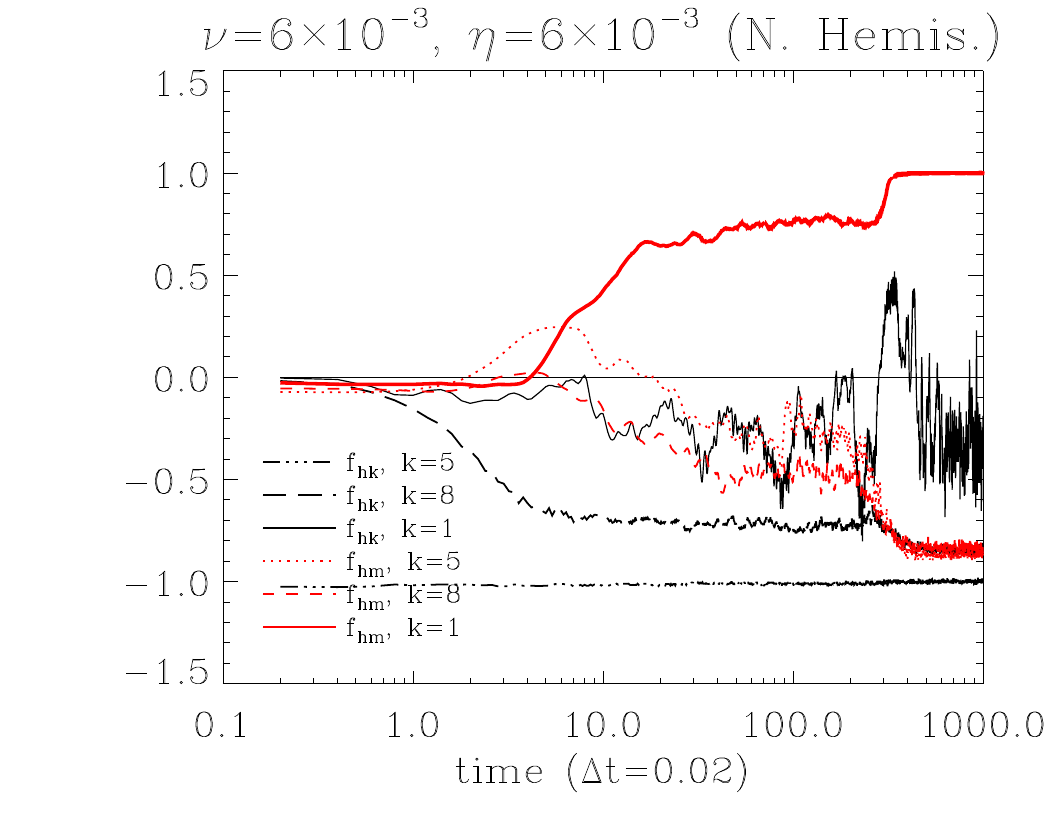}
     \label{f4d}
   }
}
\caption{(a), (b) Current helicity, defined as $H_C = k^2 \langle \mathbf{A} \cdot \bm{B} \rangle$ is used because $ \langle \mathbf{A} \cdot \bm{B} \rangle $ becomes small at $ k > 2 $.}
\end{figure*}

Figs.~\ref{f4a} and \ref{f4b} compare the current helicity $\langle \mathbf{J}\cdot \mathbf{B} \rangle (=k^2H_M)$ with the magnetic energy multiplied by the wave number $k\langle B^2 \rangle$ across the entire Fourier space. In Fig.~\ref{f4a}, which represents the Southern Hemisphere, the signs of $\langle \mathbf{J}\cdot \mathbf{B} \rangle (=k^2H_M)$ and $k\langle B^2 \rangle$ are opposite in the large-scale region where $k=1$, while they are the same on other scales. This results from the positive kinetic helicity generated by buoyancy and the leftward-deflecting Coriolis force in the Southern Hemisphere. The field structure in Fig.~2(b) corresponds to the southern hemisphere. $\langle \mathbf{j}_0\cdot \mathbf{B}_0\rangle$ corresponds to current helicity at $k=1$, but $\langle \mathbf{j}_1\cdot \mathbf{b}_{ind}\rangle$ corresponds to current helicity at $k>2$. Conversely, Fig.~\ref{f4b} shows the opposite behavior in the Northern Hemisphere.\\

Figs.~\ref{f4c} and \ref{f4d} display the kinetic helicity ratio $f_{hk}(=\langle \mathbf{U}\cdot \mathbf{\omega} \rangle/ k\langle U^2\rangle)$ (black) and the magnetic helicity ratio $f_{hm}(=k\langle \mathbf{A}\cdot \mathbf{B} \rangle/ \langle B^2\rangle$ (red) for $k=1,\, 5,\, 8$. In the Southern Hemisphere, $f_{hk}$ at the forcing scale, where $k=5$, is $+1$, and $f_{hk}$ in other regions maintains somewhat positive values. On the other hand, $f_{hm}$ converges to $-1$ when $k=1$, with values in other small-scale regions converging to positive values. The opposite phenomenon occurs in the Northern Hemisphere, as shown in the right panel.\\

Figs.~\ref{f5a} and \ref{f5b} compare the $\alpha_{\mathrm{EM-HM}}$ coefficients obtained using large-scale magnetic data [Eq.~(\ref{alphaSolution3})] with the $\alpha_{\mathrm{MFT}}$ coefficients approximated using small-scale kinetic and magnetic data [Eq.~(\ref{appen_alpha_beta_MFT}) in Appendix]. As Fig.~\ref{f5a} indicates, when the plasma system is driven with positive kinetic helicity, the $\alpha$ effect decreases from positive to negative and then converges to 0. Conversely, when the system is driven with negative kinetic helicity, as shown in Fig.~\ref{f5b}, the $\alpha$ effect maintains a positive value before converging to 0. The MFT method requires integrating residual helicity over time, given by $ \frac{1}{3} \int^{\tau} \left( \langle \mathbf{j} \cdot \mathbf{b} \rangle - \langle \mathbf{u} \cdot \mathbf{\omega} \rangle \right) \, dt $, but since the exact time range is unknown, only the residual helicity and the coefficient $\frac{1}{3}$ are considered here. We calculated $\alpha_{\mathrm{MFT}}$ values across cases of $k = 2 - 4$, $k = 2 - 6$, and $k = 2 - k_{\text{max}}$. In both hemispheres, $\alpha \sim 0$ is obtained for $k = 2 - 4$, and there is no difference in $\alpha$ between $k = 2 - 6$ and $k = 2 - k_{\text{max}}$, indicating that the forcing scale $k = 5$ primarily determines $\alpha$.\\

The discrepancy between $\alpha_{\mathrm{EM\text{-}HM}}$ and $\alpha_{\mathrm{MFT}}$ suggests that the higher-order terms neglected in the derivation of $\alpha_{\mathrm{MFT}}$ may contribute to the quenching of the $\alpha_{\mathrm{MFT}}$ effect, thereby limiting its role in magnetic field amplification. Moreover, the evolving profiles of $\alpha_{\mathrm{EM\text{-}HM}}$ indicate that the $\alpha$ effect is relatively weak in the dynamo process. This is reasonable for an electrically neutral plasma system composed of numerous charged particles. The influence of electromagnetic forces—specifically, the $\alpha$ effect—is inevitably somewhat limited, and the collective behavior of clustered particles exhibits strong fluid-like characteristics, such as diffusion, from a statistical perspective. This is also consistent with the fact that the phase difference between the magnetic fields $ B_{\mathrm{pol}} $ and $ B_{\mathrm{tor}} $ in the actual Sun is $ \pi/2 $. As $ B_{\mathrm{tor}} $ begins to decrease, $ B_{\mathrm{pol}} $ starts to arise, and vice versa. In the presence of a strong $\alpha$ effect, the interaction between the two fields is restricted to either in-phase (zero-mode) or anti-phase ($ \pi $-mode) coupling \citep{2002clme.book.....G}.\\

Figs.~\ref{f5c} and \ref{f5d} illustrate the magnetic diffusion, or $\beta$ effect. Here, we compare $\beta_{\mathrm{EM-HM}}$, obtained from large-scale magnetic data [Eq.~(\ref{betaSolution3})]; $\beta_{\mathrm{MFT}}$ (or $\beta_{vv}$), derived from plasma turbulent kinetic energy [Eq.~(\ref{appen_beta_MFT}) in Appendix]; and $\beta_{\mathrm{vv-vw}}$, calculated using plasma turbulent kinetic energy and kinetic helicity [Eq.~(\ref{general_beta_derivation8})]. The coefficient $\beta_{\mathrm{EM-HM}}$ maintains a negative value before converging to zero. Since the $\beta$ effect is always accompanied by the Laplacian operator, i.e., $\beta \nabla^2 \rightarrow -\beta k^2$, the negative $\beta$ contributes to the diffusion of the magnetic field toward larger scales and the amplification of the magnetic field strength. Similar effects are observed for $\beta_{\mathrm{EM-HM}}$ in both the Northern and Southern Hemispheres. Meanwhile, $\beta_{\mathrm{MFT}}$ remains positive, yielding a negative effect in conjunction with the Laplacian, which ultimately serves to decrease magnetic field energy. This result accounts for the leading term, without considering the helical component of plasma kinetic energy. By contrast, $\beta_{\mathrm{vv\text{-}vw}}$, which incorporates both kinetic energy and kinetic helicity, aligns closely with $\beta_{\mathrm{EM\text{-}HM}}$, suggesting that the kinetic helicity effect in the plasma plays a crucial role in the actual magnetic diffusion. Moreover, the similarity between $\beta_{\mathrm{vv\text{-}vw}}$ evaluated over the $k = 2$–6 range and that over the entire $k = 2$–$k_{\mathrm{max}}$ range indicates that the external forcing scale at $k=5$ substantially determines the diffusion.\\

In terms of observations, $\beta_{EM-HM}$, which requires information on large-scale magnetic fields, is more useful than $\beta_{vv-vw}$ from turbulent regions. However, since $\beta_{vv-vw}$ includes insights into plasma mechanical activities within the celestial plasma systems, further research into $\beta_{vv-vw}$ is necessary. The alignment of $\beta_{EM-HM}$ and $\beta_{vv-vw}$ in the $10 < t < 250$ range suggests the critical, independent role of kinetic helicity in magnetic field amplification through diffusion. However, the discrepancy observed in the region where the magnetic field undergoes significant amplification ($t>\sim 250$) implies that $\beta_{vv-vw}$ should incorporate magnetic effects.\\

Figs.~\ref{f6a} and \ref{f6b} present a comparison between $\alpha$ and $\beta$ parameters with EMF from DNS. What we need is $\langle {\bf u} \times {\bf b} \rangle$ in the turbulent region, but since the range is uncertain, we instead used $\partial \overline{\mathbf{B}}/\partial t -\eta \nabla^2\overline{\mathbf{B}}$, which corresponds to $\nabla \times \langle {\bf u} \times {\bf b} \rangle$. In contrast, $\nabla \times (\alpha \overline{\bf B} - \beta \nabla \times \overline{\bf B})$ can be directly calculated for each case. We compare ($\alpha_{EM-HM}$, $\beta_{EM-HM}$) with ($\alpha_{EM-HM}$, $\beta_{vv-vw}$), finding that the DNS result aligns closely with ($\alpha_{EM-HM}$, $\beta_{EM-HM}$).\\

Figs.~\ref{f6c} and \ref{f6d} display the large-scale magnetic fields reproduced using three different combinations of coefficients: ($\alpha_{EM-HM}$, $\beta_{EM-HM}$), ($\alpha_{EM-HM}$, $\beta_{VV-HV}$), and ($\alpha_{MFT}$, $\beta_{MFT}$). These results are compared against the direct numerical simulation (DNS) results to assess the accuracy and efficacy of each coefficient pairing. We used an IDL script as follows to reproduce the magnetic field.
\begin{verbatim}
B[0] =  sqrt(2.0*spec_mag(1, 0))  % k=1 for large scale

for j=0L,  t_last do begin
  B[j+1] = B[j] + (sign*alpha[j]-beta[j]-eta)*B[j]*(time[j+1]-time[j])
  % helical magnetic field
endfor
\end{verbatim}
Here, the time interval $\Delta t \sim 0.2$ and the correlation length $l = 2\pi / 3$ are used. The sign in front of $\alpha$ is positive (`$+$') for the Northern Hemisphere (left-handed kinetic helicity, Figs~\ref{f4b}, \ref{f4d}, $k=1,\,5$) and negative (`$-$') for the Southern Hemisphere (right-handed kinetic helicity, Figs~\ref{f4a}, \ref{f4c}, $k=1,\,5$). Among the configurations, ($\alpha_{EM-HM}$ \& $\beta_{EM-HM}$) yields the most accurate results, while the classical MFT method shows the lowest accuracy. This trend is consistent with the results observed in Fig.~5. Meanwhile, ($\alpha_{EM-HM}$ \& $\beta_{vv-vw}$) provides relatively accurate results in the kinematic regime. However, as the magnetic field grows, $\beta_{vv-vw}$ deviates larger than the actual value. This unconstrained growth implies the existence of additional mechanisms that quench the $\beta$ effect.\\

The $\beta$ effect including the magnetic effect is shown in Fig.~\ref{f7a}---namely, the magnetic energy $ E_M \sim \langle b^2\rangle $ and current helicity $ H_C \sim \langle\mathbf{j} \cdot \mathbf{b}\rangle $. Compared to Fig.~\ref{f5c}, there is little difference in $\beta$ in the kinematic regime. However, in the nonlinear regime, the new $\beta$ profile (red solid line) efficiently decreases toward zero. Based on this, the reproduced magnetic field is shown in Fig.~\ref{f7b}. Compared to Fig.~\ref{f6c}, the new magnetic field remains close to the DNS results. Here, the magnetic energy $\langle b^2\rangle$ in the turbulent region plays a similar role to $\langle v^2\rangle$, whereas the current helicity $\langle\mathbf{j} \cdot \mathbf{b} \rangle$ acts oppositely to the kinetic helicity $\langle\mathbf{v} \cdot \boldsymbol{\omega}\rangle$, thereby reducing the overall $\beta$.\\





Figs.~\ref{f8a} and \ref{f8b} show how the $\alpha$ and $\beta$ effects contribute to the evolution of large-scale magnetic fields. In Fig.~\ref{f8a}, the $\beta_{\mathrm{EM\text{-}HM}}$ effect is fully applied, while the $\alpha_{\mathrm{EM\text{-}HM}}$ effect is modified. The dot-dashed line represents the case where both $\alpha$ and $\beta$ effects are fully included. The dot-dot-dot-dashed line shows the result with only the $\beta$ effect applied, and the red dashed line shows the case where the $\alpha$ effect is applied with the opposite sign. This figure shows that the $\alpha$ effect is not important during the early (kinematic) stage, but it becomes more significant as the magnetic field grows and reaches saturation. On the other hand, in Fig.~\ref{f8b}, the $\alpha_{\mathrm{EM\text{-}HM}}$ effect is fully applied, and the $\beta_{\mathrm{EM\text{-}HM}}$ effect is reduced. The dot-dot-dot-dashed line shows the result with 50\% of the $\beta$ effect, and the red dashed line shows the case with only 15\% of it. When the $\beta$ contribution drops below around 10\%–15\%, the magnetic field stops growing. In many previous dynamo theories, the $\alpha$ effect was considered the main factor in building large-scale magnetic fields. However, these results repeatedly demonstrate that the $\beta$ effect plays a more important role in the amplification of the large-scale magnetic field.\\



\section{Theoretical Framework}

\subsection{Derivation of $\alpha$ and $\beta$ Coefficients}
There have been numerous theoretical attempts to derive the $\alpha$ and $\beta$ coefficients. Among them, the important ones are included in the appendix, while here we primarily present our model.\\


From Eq.~(\ref{magnetic induction equation_alpha_beta}), we constructed the coupled equations for $\overline{H}_M(t)$ and $\overline{E}_M(t)$ as follows \citep{2020ApJ...898..112P, 2023ApJ...944....2P, 2025PhRvD.111b3021P}:
\begin{eqnarray}
\frac{\partial \overline{H}_M}{\partial t} &=&  4\alpha \overline{E}_M - 2(\beta + \eta)\overline{H}_M, \label{Hm1} \\
\frac{\partial \overline{E}_M}{\partial t} &=&  \alpha \overline{H}_M - 2(\beta + \eta)\overline{E}_M. \label{Em1}
\end{eqnarray}
For the large scale field with $k=1$, magnetic helicity and current helicity coincide: $\langle \mathbf{J}\cdot \mathbf{B}\rangle = k^2\langle \mathbf{A}\cdot \mathbf{B}\rangle\rightarrow \langle \mathbf{A}\cdot \mathbf{B}\rangle$\footnote{Magnetic helicity is defined as the inner product of an axial vector $\bf B$ and a polar vector $\bf A$ to be a pseudoscalar. $\alpha$ is also considered as a pseudoscalar while $\overline{E}_M$ and $\beta$ are normal scalars.}.\\

The solutions are
\begin{eqnarray}
2\overline{H}_M(t)&=&(2\overline{E}_{M0}+\overline{H}_{M0})e^{2\int^{t}_0(\alpha-\beta-\eta)d\tau}-(2\overline{E}_{M0}-\overline{H}_{M0})e^{2\int^{t}_0(-\alpha-\beta-\eta)d\tau},\label{HmSolutionwithAlphaBeta1}\\
4\overline{E}_{M}(t)&=&(2\overline{E}_{M0}+\overline{H}_{M0})e^{2\int^{t}_0(\alpha-\beta-\eta)d\tau}+(2\overline{E}_{M0}-\overline{H}_{M0})e^{2\int^{t}_0(-\alpha-\beta-\eta)d\tau},\label{EmSolutionwithAlphaBeta2}
\end{eqnarray}
and, $\alpha$ and $\beta$ are
\begin{eqnarray}
\alpha_{EM-HM}&=&\frac{1}{4}\frac{d}{dt}log_e \bigg|\frac{ 2\overline{E}_M(t)+\overline{H}_M(t)}{2\overline{E}_M(t)-\overline{H}_M(t)}\bigg|,\label{alphaSolution3}\\
\beta_{EM-HM}&=&-\frac{1}{4}\frac{d}{dt}log_e\big| \big(2\overline{E}_M(t)-\overline{H}_M(t) \big)\big( 2\overline{E}_M(t)+\overline{H}_M(t)\big)\big|-\eta.\label{betaSolution3}
\end{eqnarray}
Substituting this result into Eq.~(\ref{Hm1}) and Eq.~(\ref{Em1}) confirms equality between the left and right sides. Note that $\alpha$ and $\beta$ are functions of large-scale or mean magnetic data expressed in differential form rather than integral form.\\

If we apply a reflection symmetry, $\overline{E}_M$ does not change sign, but $\overline{H}_M$ does. Then, the denominator and numerator in Eq.~(\ref{alphaSolution3}) are reversed, leading to a change in the sign of $\alpha_{EM-HM}$; that is, it behaves as a pseudoscalar. In contrast, $\beta_{EM-HM}$ in Eq.~(\ref{betaSolution3}) remains unchanged under reflection and thus is a scalar. For reference, helicities such as $\mathbf{A} \cdot \mathbf{B}$, $\mathbf{U} \cdot \nabla\times \mathbf{U}$, and $\mathbf{U} \cdot \mathbf{B}$ represent interactions between an axial vector and a polar vector. Moreover, the electromagnetic field tensor $F_{\mu\nu}$ combines the electric field (a polar vector) and the magnetic field (an axial vector) into a single relativistically consistent object. Their different transformation properties are properly handled by the antisymmetric tensor structure. It is fundamentally incorrect to reject the validity of an equation or a physical quantity merely because it involves combinations of vectors and scalars of different transformation properties.\\



\subsection{Derivation of $\beta$ using turbulent kinetic data}
With Eqs.~(\ref{alphaSolution3}) and (\ref{betaSolution3}), the profiles of $\alpha$ and $\beta$ can be determined exactly. However, this indirect approach does not explain the physical mechanisms by which these effects are formed. In this section, we derive a more general $\beta$ again using the function iterative approach, with a more detailed statistical identity relation.\\

Conventionally, $\beta$ is found with
\begin{eqnarray}
EMF=\bigg\langle {\bf u}\times \int^{\tau}(-{\bf u}\cdot \nabla \overline{\bf B})dt \bigg\rangle \sim \bigg\langle -\epsilon_{ijk} u_j (r)u_m(r+l)\tau \frac{\partial \overline{B}_k}{\partial \overline{r}_m}\bigg\rangle \rightarrow -\frac{\tau}{3}\langle u^2\rangle\epsilon_{ijk}\frac{\partial \overline{B}_k}{\partial \overline{r}_m}\delta_{jm}. \label{beta_derivation_helical_first}
\end{eqnarray}
The eddy turnover time $\tau$ can be set as 1 under the assumption that the two eddies $u_j$ and $u_m$ are correlated over one eddy turnover time. However, assuming the spatial correlation length `$ l $' for $ u_j $ and $ u_m $ to be $ l \rightarrow 0 $ and replacing the second-order velocity moment with kinetic energy is overly simplified. The turbulent magnetic fluctuation $\mathbf{b}$ in EMF arises from the cross-correlation between $\mathbf{u}$ and the mean magnetic field $\overline{\mathbf{B}}$. During this process, temporal and spatial causal relationships are established between them. Therefore, when considering $\mathbf{u}$ and $\mathbf{b}$, the two-point cross correlation $\langle \mathbf{u}(\mathbf{r}) \times \mathbf{b}(\mathbf{r} + \mathbf{l}) \rangle$ must be used. The formal expression of EMF, $\xi_i = \alpha_{ij} \overline{B}_j- \beta_{ijk}\partial \overline{B}_j/ \partial x_k... $ inherently assumes such two-point correlations.\\

\begin{figure*}
    {
   \subfigure[]{
     \includegraphics[width=8.3 cm]{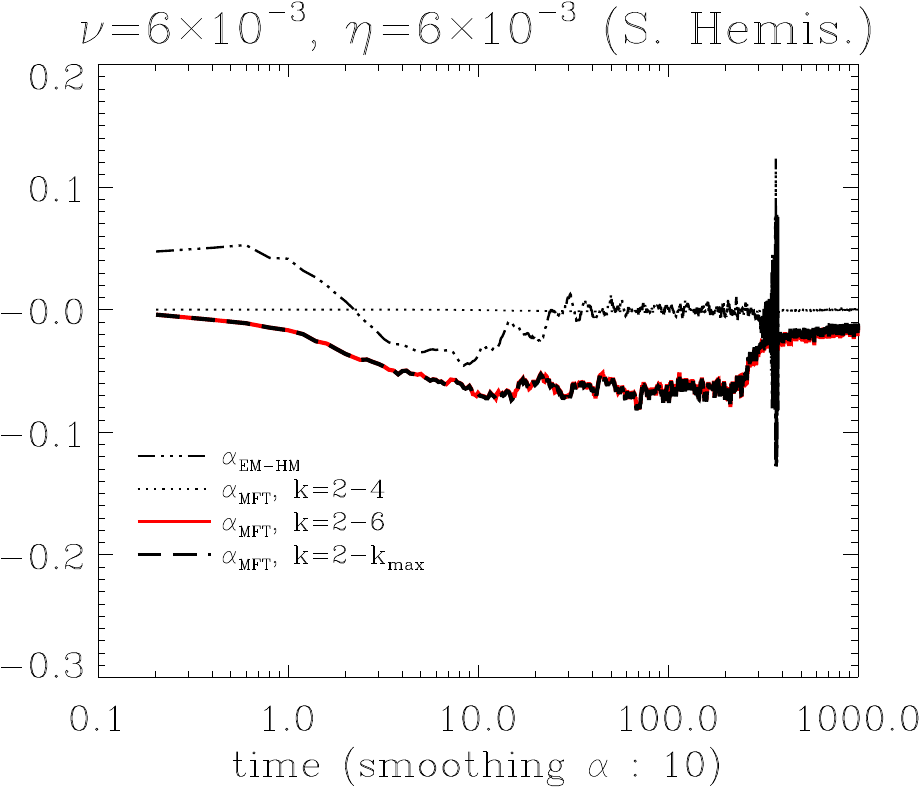}
     \label{f5a}
    }\hspace{-13 mm}
   \subfigure[]{
   \includegraphics[width=8.3 cm]{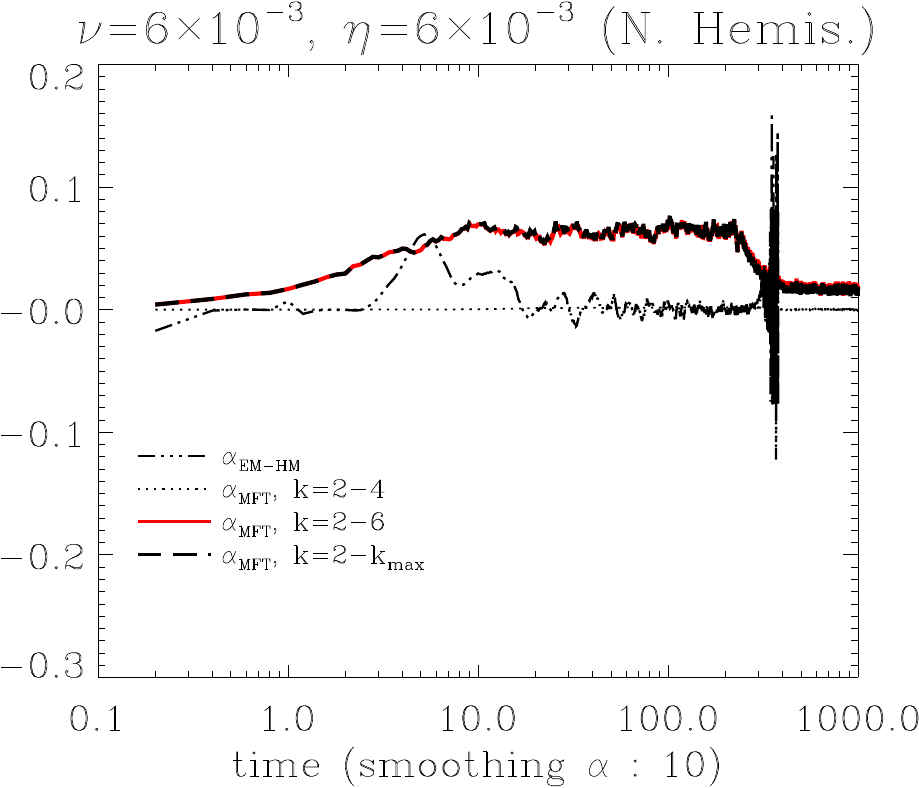}
     \label{f5b}
   }\hspace{-13 mm}
   \subfigure[]{
   \includegraphics[width=8.5 cm]{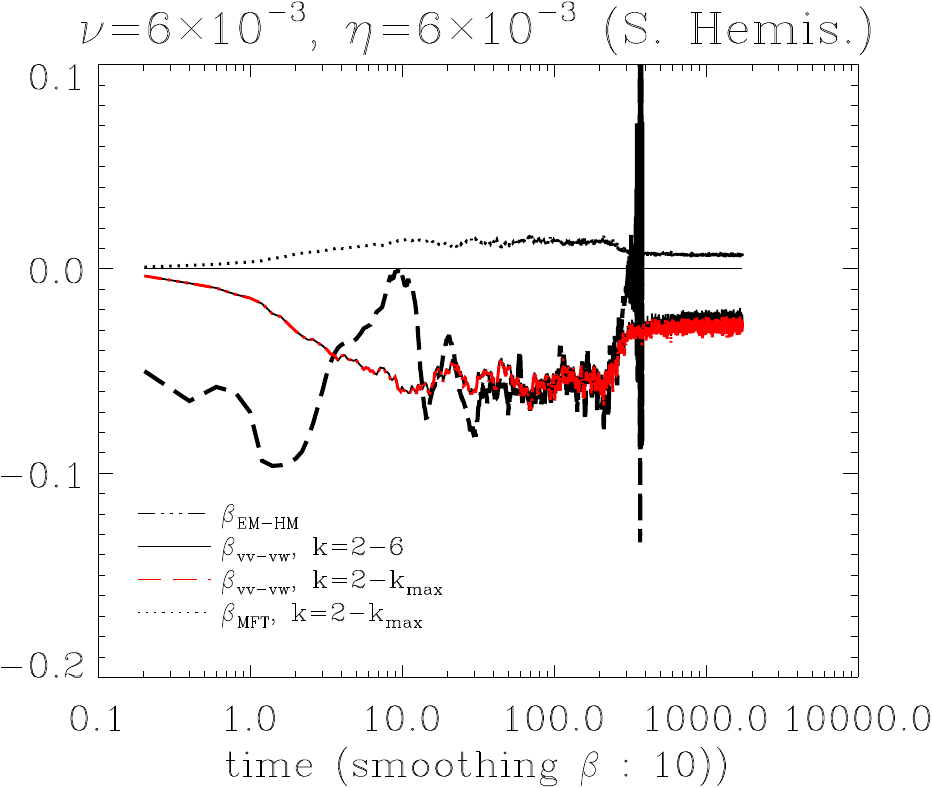}
     \label{f5c}
   }\hspace{6 mm}
   \subfigure[]{
   \includegraphics[width=8.5 cm]{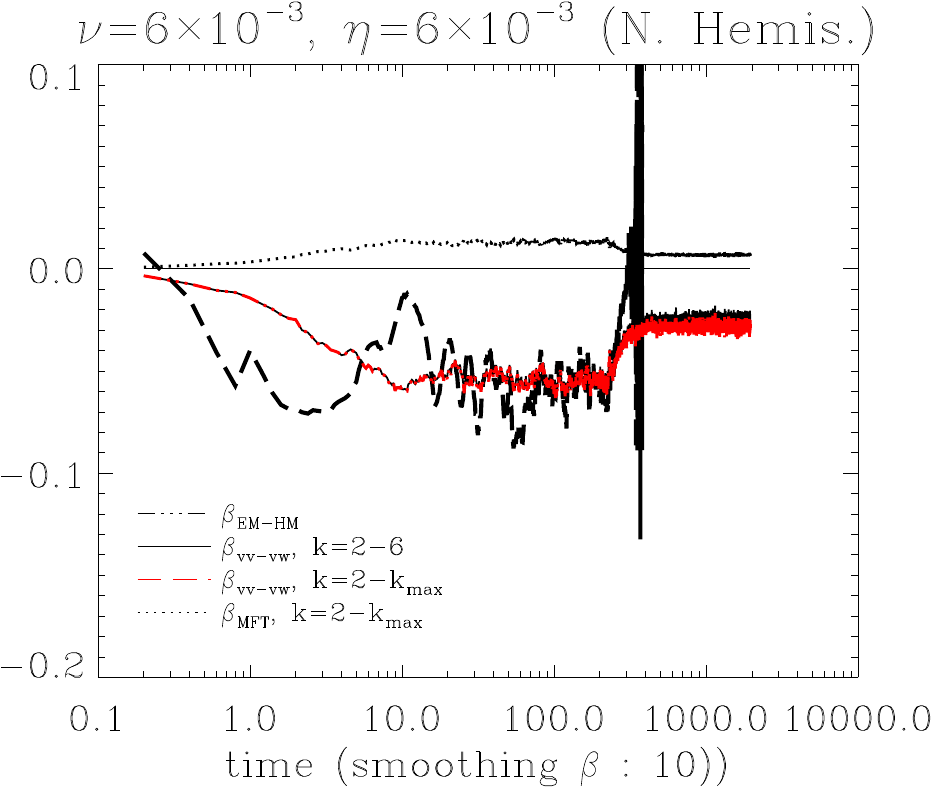}
     \label{f5d}
   }
}
\caption{(a), (b) $\alpha_{MFT}(\rightarrow \alpha_{jb-vw})$ reflects its dependence on the sign of externally provided kinetic helicity. (c), (d) Unlike $\alpha$, the $\beta$ coefficient does not change its sign in response to the sign of kinetic helicity.  $\beta_{EM-HM}$ exhibits significant fluctuations after $t > 250$, but its time average effectively converges to zero. In contrast, $\beta_{MFT}$ (i.e., $\beta_{vv}$) remains positive, while $\beta_{vv-vw}$ stays negative.}

\end{figure*}

\begin{figure*}
    {
   \subfigure[]{
     \includegraphics[width=9.4 cm]{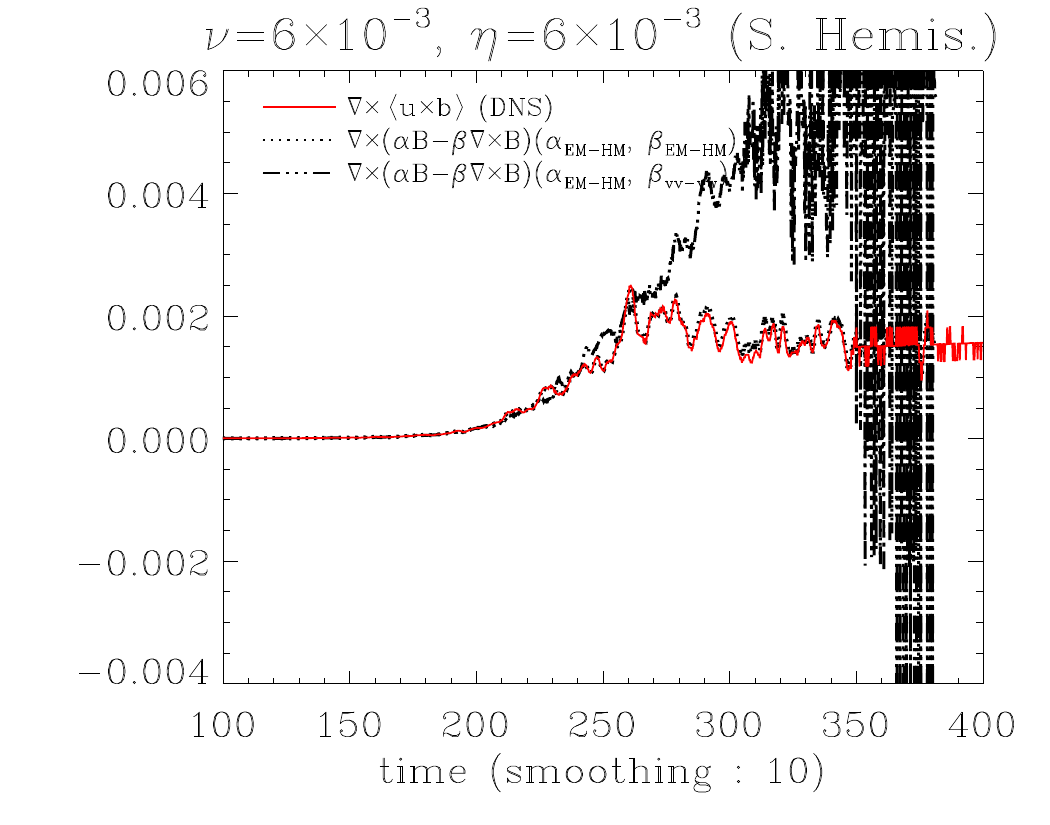}
     \label{f6a}
    }\hspace{-6 mm}
   \subfigure[]{
   \includegraphics[width=9.4 cm]{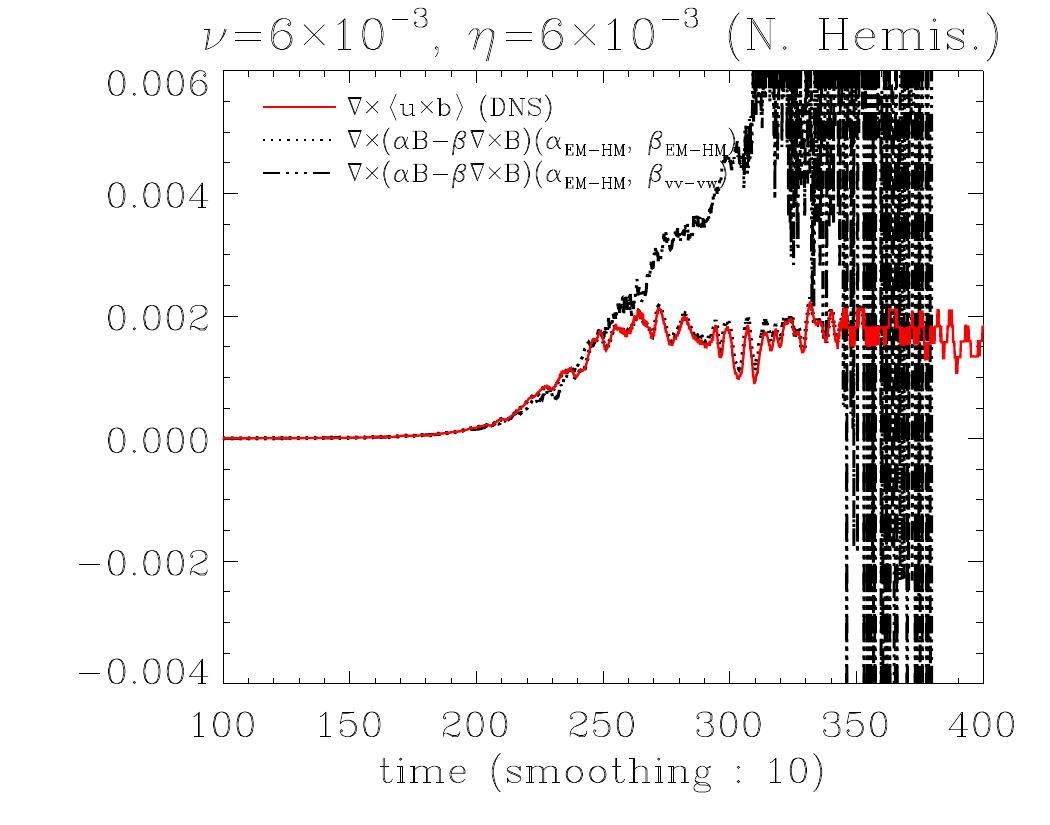}
     \label{f6b}
   }\hspace{-13 mm}
   \subfigure[]{
   \includegraphics[width=9.4 cm]{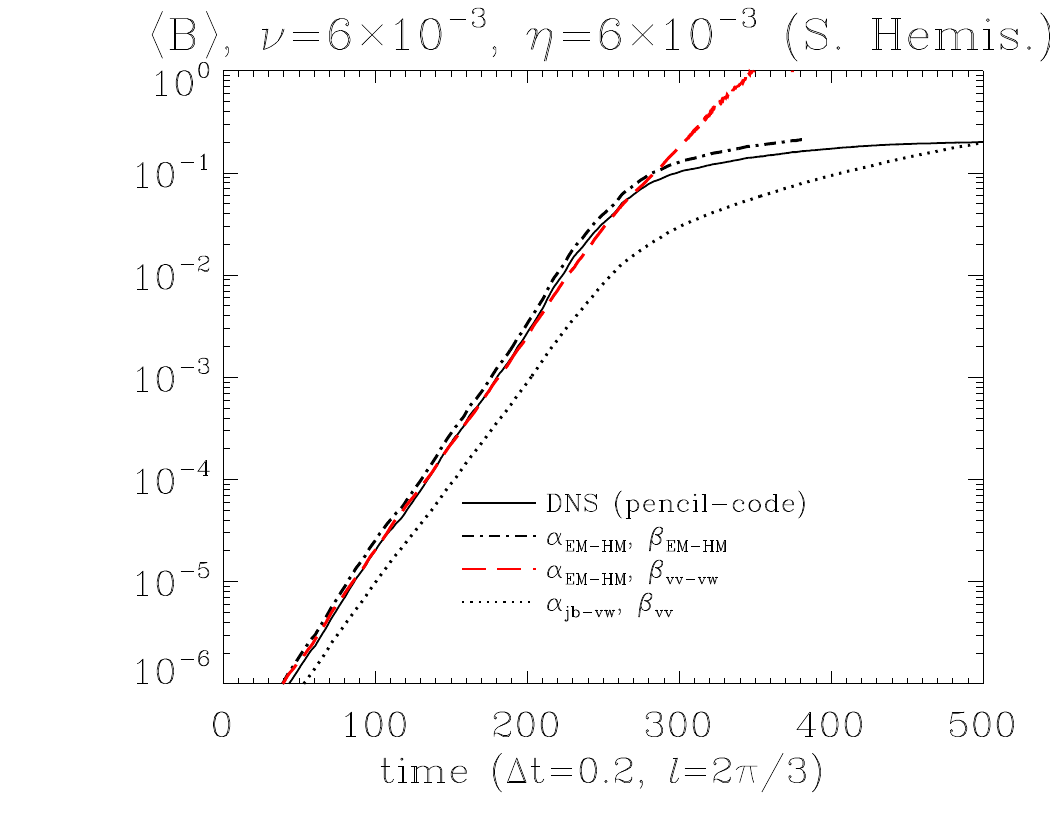}
     \label{f6c}
   }\hspace{-6 mm}
   \subfigure[]{
   \includegraphics[width=9.4 cm]{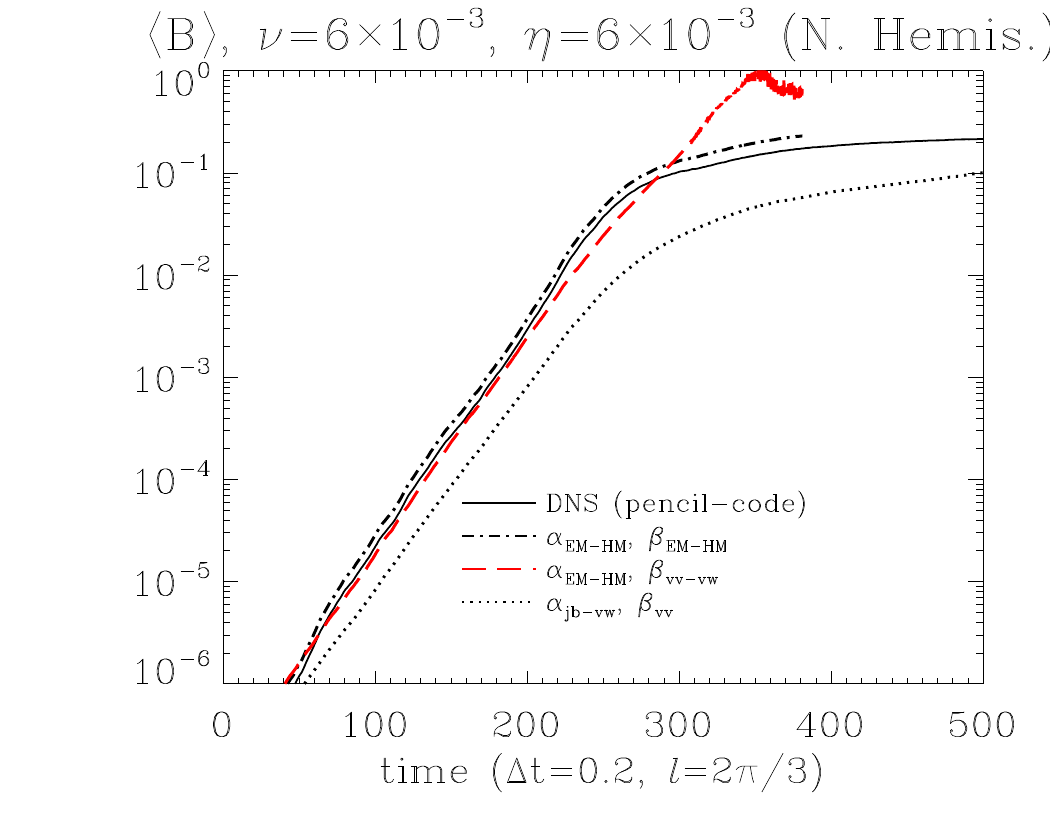}
     \label{f6d}
   }
}
\caption{(a), (b) Comparison of the EMF reconstructed by substituting $\alpha$ and $\beta$. The values of $\alpha_{\mathrm{EM-HM}}$ and $\beta_{\mathrm{EM-HM}}$ show good agreement with the directly measured EMF, $\langle \mathbf{u} \times \mathbf{b} \rangle$. In contrast, $\beta_{\mathrm{VV-VW}}$ deviates after $t \sim 250$.
(c), (d) Large-scale magnetic field $\overline{B}$ reconstructed using different combinations of $\alpha$ and $\beta$.}
\end{figure*}

In our previous work, we derived an expression for the $\beta$ coefficient that includes both turbulent kinetic energy and kinetic helicity, using a more general second-order moment identity that accounts for helicity effects.
\begin{eqnarray}
\bigg\langle \mathbf{u}\times \int^{\tau}\frac{\partial \mathbf{b}}{\partial t}dt \bigg\rangle&\rightarrow&
\bigg\langle -\epsilon_{ijk} u_j (r)u_m(r+l)\frac{\partial \overline{B}_k}{\partial \overline{r}_m}\bigg\rangle\nonumber\\
&\sim&\bigg(\frac{1}{3}\langle u^2\rangle - \frac{l}{6}H_V \bigg)\,\big(-\nabla \times \overline{\bf B}\big)\equiv \beta_{vv-vw} \, \big(-\nabla \times \overline{\bf B}\big).\label{general_beta_derivation8}
\end{eqnarray}
($\bm{l}=\mathbf{r}_1-\mathbf{r}$ is a separation vector.) This expression accurately reproduces the results when the magnetic field is weak. However, it has a limitation in the nonlinear regime where the magnetic field becomes strong, as it fails to capture the quenching effect adequately (see Fig.~\ref{f5c}). Therefore, $\beta$ needs to be derived in a more general form,
\begin{eqnarray}
EMF =  \int^{\tau}\frac{\partial}{\partial t}\big\langle \mathbf{u} \times \mathbf{b} \rangle\,dt= \bigg\langle \int^{\tau}\frac{\partial \mathbf{u}}{\partial t}dt \times \mathbf{b} \bigg\rangle +
\bigg\langle \mathbf{u}\times \int^{\tau}\frac{\partial \mathbf{b}}{\partial t}dt \bigg\rangle\label{more_general_beta_derivation1}
\end{eqnarray}
For the first term in RHS,
\begin{eqnarray}
\bigg\langle \int^{\tau}\frac{\partial \mathbf{u}}{\partial t}dt \times \mathbf{b} \bigg\rangle &\rightarrow&
\tau\big\langle \mathbf{b}\cdot\nabla \overline{\mathbf{B}} \times \mathbf{b} \big\rangle\sim \tau\bigg\langle \epsilon_{ijk} b_m\frac{\partial \overline{B}_j}{\partial \overline{r}_m}b_k\bigg\rangle\nonumber\\
&\sim&\frac{\tau}{3}\langle b_kb_m\rangle \epsilon_{ijk}\frac{\partial \overline{B}_j}{\partial \overline{r}_m}\delta_{km}-
\tau\big\langle \epsilon_{ijk}\epsilon_{kms}\frac{l_s}{6}\big(\mathbf{b}\cdot \nabla\times \mathbf{b}\big)\big\rangle\frac{\partial \overline{B}_j}{\partial \overline{r}_m}\,\,(m\rightarrow i,\,s\rightarrow j)\nonumber\\
%
%
&\rightarrow&-\frac{\tau}{3}\langle b^2\rangle \big(\nabla \times \overline{\bf B}\big)_i-\tau \frac{H_c}{6}\big(\bm{l}\times \big[\nabla\times \overline{\bf B}\big]\big)_i\,\,\,\nonumber\\
&\rightarrow&-\frac{\tau}{3}\langle b^2\rangle \big(\nabla \times \overline{\bf B}\big) - \tau\frac{l}{6}H_C \big(\nabla \times \overline{\bf B}\big)\nonumber\\
&\rightarrow&\tau\bigg(\frac{1}{3}\langle b^2\rangle + \frac{l}{6}H_C \bigg)\,\big(-\nabla \times \overline{\bf B}\big)\equiv \beta_{bb+jb} \, \big(-\nabla \times \overline{\bf B}\big).\label{general_beta_derivation_for_b8}
\end{eqnarray}
We have used the identity of a second order moment for magnetic field with current helicity $H_C=\langle \mathbf{j}\cdot \mathbf{b}\rangle$ (see Eq.~(\ref{appen_general_beta_derivation5}) in Appendix):
\begin{eqnarray}
\langle b_k(r)b_m(r+l)\rangle=\frac{\langle b^2\rangle}{3}\delta_{km}-\epsilon_{kms}\frac{l_s}{6}H_C.
\label{general_beta_derivation5}
\end{eqnarray}

\begin{figure*}
    {
   \subfigure[]{
   \includegraphics[width=9 cm]{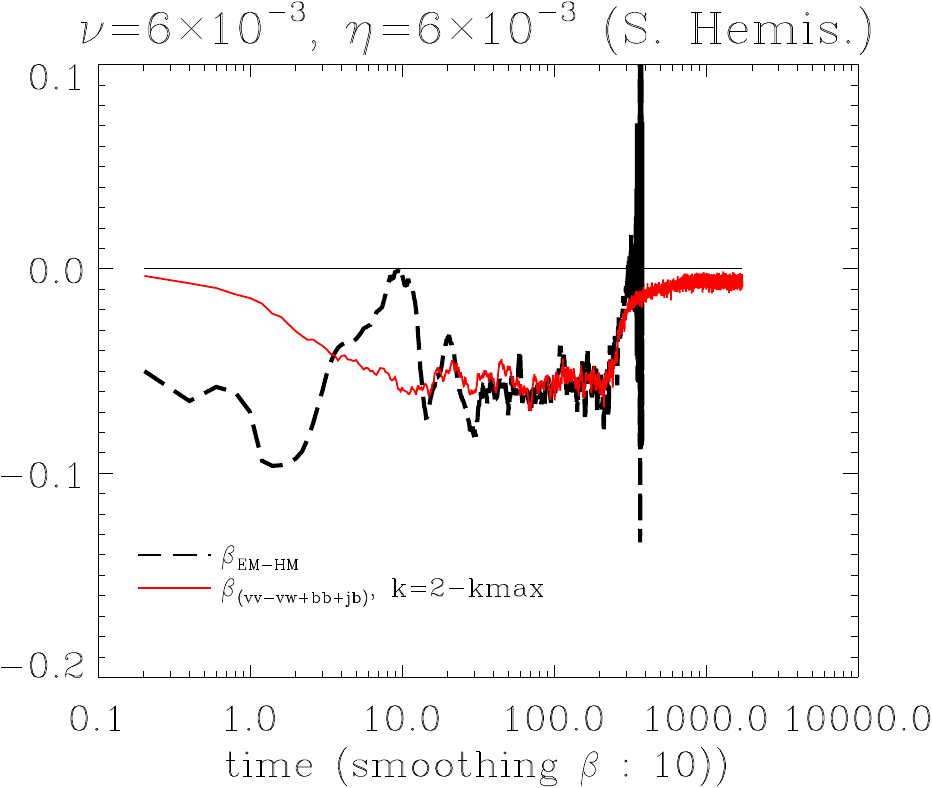}
     \label{f7a}
   }\hspace{2 mm}
   \subfigure[]{
   \includegraphics[width=8.4 cm]{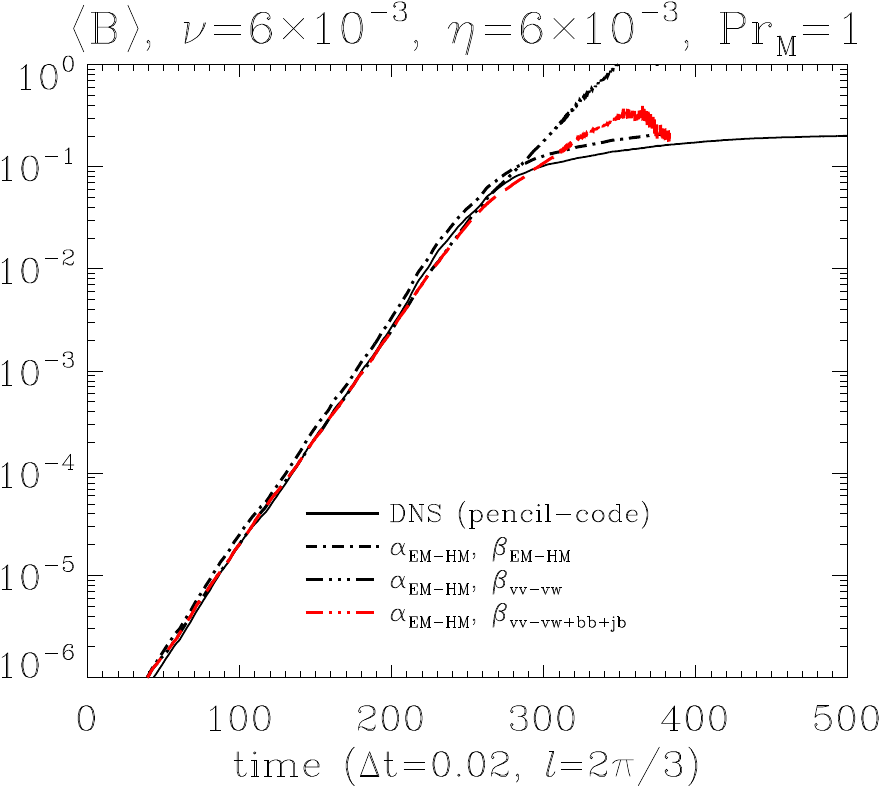}
     \label{f7b}
   }
}
\caption{(a) $\beta$ derived from turbulent velocity and magnetic fields (see Eq.~(\ref{complete_beta_derivation}).
(b) New large-scale magnetic fields including both turbulent kinetic and magnetic effects (red dot-dot-dot-dashed line).}
\end{figure*}

\begin{figure*}
    {
   \subfigure[]{
   \includegraphics[width=8.6 cm]{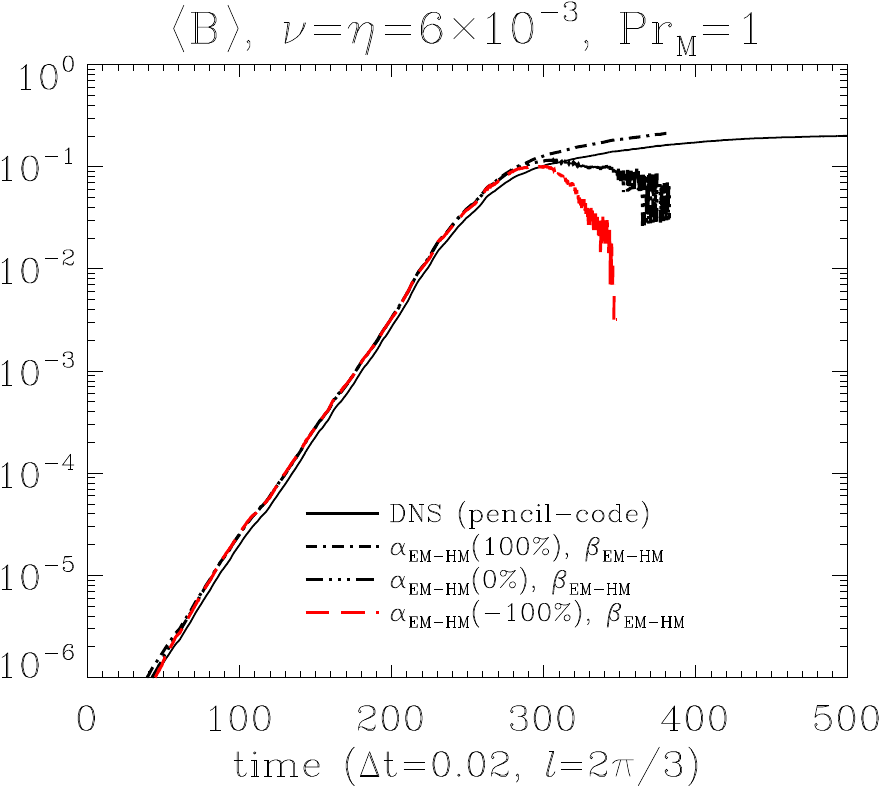}
     \label{f8a}
   }\hspace{6 mm}
   \subfigure[]{
   \includegraphics[width=8.6 cm]{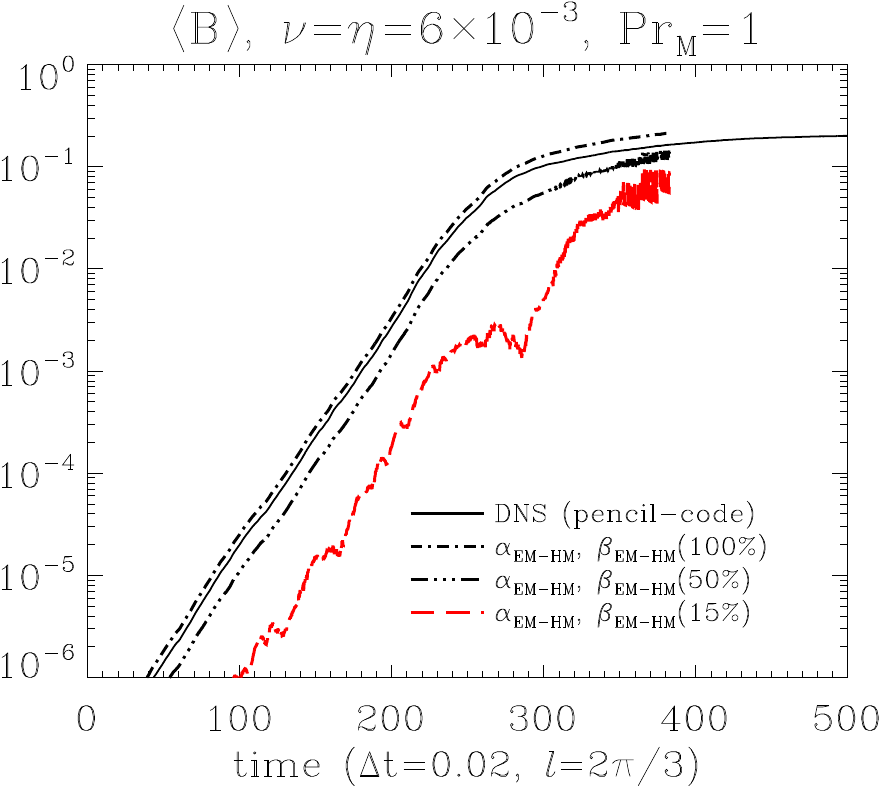}
     \label{f8b}
   }
}
\caption{(a) Evolution of large-scale magnetic fields reconstructed by varying the weight of $\alpha$ from $+100$\% to $-100$\%, while keeping the full $\beta$ unchanged. (b) $\overline{B}$ field evolution with varying $\beta$ from $+100$\% to $15$\%, under the full $\alpha$ effect. ($\overline{B}\rightarrow 0$ under $\beta<$$15$\%)}
\end{figure*}

The electromotive force (EMF), given by $\langle \mathbf{u} \times \mathbf{b} \rangle$, is formally a polar vector. The term $\bm{l} \times (\nabla \times \overline{\mathbf{B}})$ is an axial vector, since it is the cross product of two polar vectors. The current helicity $H_C = \langle \mathbf{j} \cdot \mathbf{b} \rangle$ is a pseudoscalar. The full expression represents a polar vector. Therefore, the more complete form of the $\beta$ coefficient can be expressed as
\begin{eqnarray}
\beta_{net} = \beta_{vv-vw}+\beta_{bb+jb} = \int^{\tau}\bigg(\frac{2}{3}E_V + \frac{2}{3}E_M - \frac{l}{6}H_V + \frac{l}{6}H_C\bigg) dt.
\label{complete_beta_derivation}
\end{eqnarray}
This expression shows that the total energy and current helicity contribute to the diffusion of the large-scale magnetic field, whereas the kinetic helicity amplifies it. This result is notably consistent with the $\alpha$ effect, in which kinetic helicity enhances the magnetic field while current helicity suppresses it. We will see that helicities influence the evolution of the magnetic field not only through their interaction with current density but also through diffusion processes (Figs.~\ref{f7a}, \ref{f7b}).\\

Finally, one may wonder whether the $\alpha$ and $\beta$ coefficients obtained in this study can reproduce the solar magnetic field pattern and its 22-year cycle. In principle, if DNS or observation data for the Sun—accounting for its kinematic viscosity, magnetic diffusivity, and rotational effects—are available, one can extract $\alpha$ and $\beta$ and use Eqs.~(\ref{magnetic induction equation_poloidal}) and (\ref{magnetic induction equation_toroidal}) to reconstruct the magnetic field pattern. However, since our DNS data are based on different values of kinematic viscosity and magnetic diffusivity ($\nu=\eta=0.006$), the resulting magnetic field pattern would not match actual solar observations.\\

In contrast, \cite{2008A&A...483..949J} adopted Eqs.~(\ref{magnetic induction equation_poloidal}) and (\ref{magnetic induction equation_toroidal}) along with semi-empirically estimated $\alpha$ and $\eta$ profiles to reconstruct the solar magnetic pattern. Similarly, \cite{2013ApJ...768...16S} extracted $\alpha$ from the simulation results of \cite{2011ApJ...735...46R} and constructed an arbitrary functional form for the diffusivity. By tuning the typical values of $\alpha_0$, $\eta_0$, and $\Omega_0$, they were able to reproduce a pattern consistent with solar observations. However, it is difficult to determine the physical significance of such parameter-fitting approaches. This is essentially different from validating or refining a model by directly comparing the reconstructed magnetic field to DNS results, as shown in Figs.~6 and 7.\\

The model used in this paper, namely Eqs.~(\ref{alphaSolution3}) and (\ref{betaSolution3}), is based on the spatial derivatives of large-scale magnetic field data. This approach does not rely on ambiguous assumptions or trial-and-error procedures, making it a clear and practical method. In the case of Eq.~(\ref{complete_beta_derivation}), information on the eddy turnover time $\tau$ and the relative position vector $\bm{l}$ is required. However, these quantities are the research subjects in statistical mechanics and turbulence model, and their behavior in astrophysical contexts is itself a valuable topic worthy of further investigation.\\


\begin{table}[H]
    \centering
    \caption{Comparison of Methods for Calculating $\alpha$ and $\beta$ and Their Impact on $\overline{B}$}
    \label{table_1}
    \renewcommand{\arraystretch}{1.3}
    \begin{tabular}{|l|l|l|l|}
        \hline
        \textbf{Method} & \textbf{Data Used} & \textbf{Calculation Method} & \textbf{Accuracy of $\overline{B}$} \\
        \hline\hline
        $\alpha_{MFT}$ \& $\beta_{MFT}$ & $v$ and $b$ & integral & inaccurate \\
        \hline
        $\alpha_{EM-HM}$ \& $\beta_{EM-HM}$ & $\overline{B}$ & differentiation & accurate \\
        \hline
        $\alpha_{EM-HM}$ \& $\beta_{vv-vw}$ & $v$ and $\overline{B}$ & integral and differentiation & accurate for weak $\overline{B}$ \\
        \hline
        $\alpha_{EM-HM}$ \& $\beta_{vv-vw+bb+jb}$ & $v$, $b$, and $\overline{B}$ & integral and differentiation & reasonably accurate in nonlinear regime\\ \hline

    \end{tabular}
\end{table}

\section{Summary}
We calculated the $\alpha$ and $\beta$ effects generated by kinetic helicity, which has opposite signs in the southern and northern hemispheres of a rotating astrophysical plasma system such as the Sun. We reconstructed the large-scale magnetic field and compared it with the DNS results. The $\alpha$ coefficients in the two hemispheres exhibit opposite signs, inducing current densities that are either parallel or antiparallel to the magnetic flux. Interestingly, the direction of the current density remains unchanged relative to the magnetic flux, which undergoes periodic reversals. As a result, positive magnetic helicity is generated at large scales in the northern hemisphere, while negative magnetic helicity arises in the southern hemisphere.\\

The large-scale magnetic field $\overline{\mathbf{B}}$ reconstructed using the coefficients $\alpha_{\mathrm{EM-HM}}$ and $\beta_{\mathrm{EM-HM}}$ showed excellent agreement with the results from direct numerical simulations (DNS). In contrast, $\beta_{\mathrm{vv-vw}}$ yielded unreliable results in the nonlinear regime, indicating the limitations of conventional $\beta$ estimations based solely on the velocity field. To address this, we introduced additional turbulent correlation terms involving the magnetic field and current density, namely $\beta_{\mathrm{bb+jb}}$, and incorporated them to construct an improved coefficient, $\beta_{\mathrm{vv-vw+bb+jb}}$. This enhanced coefficient more accurately reproduced the stable and physically realistic structure of the magnetic field.\\

Furthermore, an analysis of the relative contributions of each coefficient, weighted by their respective amplitudes, revealed that, contrary to prior theoretical expectations, the $\beta$ effect plays a more dominant role than the $\alpha$ effect in the dynamo process. In particular, the electromagnetic $\alpha$ effect is found to be negligible during the kinematic phase but begins to contribute to the maintenance of the saturated magnetic field as the field grows. These findings suggest the need to reconsider the conventional understanding of the relative roles of the $\alpha$ and $\beta$ effects and highlight the importance of a more precise analysis of the $\beta$ diffusivity for a deeper understanding of the structure and evolution of turbulent magnetic fields.\\

The methods used to compute $\alpha$ and $\beta$, along with a comparison of the resulting magnetic fields, are summarized in Table~\ref{table_1}.\\

\subsection{Derivation of $\alpha$ \& $\beta$}
Several attempts have been made to calculate these coefficients using various dynamo theories such as MFT, EDQNM, and DIA. Despite these efforts, only approximations of the $\alpha$ and $\beta$ coefficients are available. These theories suggest that $\alpha$ is related to residual helicity, $\langle \mathbf{b} \cdot (\nabla \times \mathbf{b}) \rangle - \langle \mathbf{u} \cdot (\nabla \times \mathbf{u}) \rangle$, while $\beta$ is linked to turbulent energy, such as $\langle u^2 \rangle$ or $\langle b^2 \rangle$. Conventionally, $\alpha$ has been understood as a generator of magnetic fields, whereas $\beta$, in combination with molecular resistivity $\eta$, was thought to simply diffuse them. The theoretical expressions for $\alpha$ and $\beta$ are outlined below.\\\\
(i) MFT \citep{1980opp..bookR....K, 2012MNRAS.419..913P}:\\
\begin{eqnarray}
\alpha_{MFT}(\mathrm{or\,\,}\alpha_{jb-vw}) &=& \frac{1}{3}\int^{\tau} \left(\langle {\bf j}\cdot {\bf b}\rangle - \langle {\bf u}\cdot \nabla\times {\bf u}\rangle\right)\, dt, \label{appen_alpha_beta_MFT}\\
\beta_{MFT}(\mathrm{or\,\,}\beta_{vv}) &=& \frac{1}{3}\int^{\tau} \langle u^2\rangle\, dt, \label{appen_beta_MFT}
\end{eqnarray}
where $\tau$ is the correlation time, whose exact value remains unknown, isotropy is assumed, and higher-order terms are neglected.\\

\noindent (ii) DIA \cite{Akira2011}:\\
\begin{eqnarray}
\alpha_{DIA} &=& \frac{1}{3} \int d{\bf k} \int^t G \left( \langle {\bf j}\cdot {\bf b} \rangle - \langle {\bf u}\cdot \nabla\times {\bf u} \rangle \right) d\tau, \label{appen_DIA_alpha} \\
\beta_{DIA} &=& \frac{1}{3} \int d{\bf k} \int^t G \left( \langle u^2 \rangle + \langle b^2 \rangle \right) d\tau, \label{appen_DIA_beta} \\
\gamma_{DIA} &=& \frac{1}{3} \int d{\bf k} \int^t G \langle {\bf u}\cdot {\bf b} \rangle \, d\tau, \label{appen_DIA_gamma}
\end{eqnarray}
Compared to MFD, the electromotive force (EMF) is characterized by the coefficients $\alpha$, $\beta$, and $\gamma$, where $\gamma$ represents cross helicity $\langle {\bf u}\cdot {\bf b} \rangle$. Additionally, $\beta$ consists of contributions from both turbulent kinetic energy and magnetic energy. It should be noted that the dependence of $\beta$ on magnetic energy cannot be derived through the function recursion approach.\\

\noindent (iii) EDQNM \citep{1975JFM....68..769F, 1976JFM....77..321P}:\\
\begin{eqnarray}
\alpha_{QN} &=& \frac{2}{3}\int^{t} \Theta_{kpq}(t)\left(\langle {\bf j}\cdot {\bf b} \rangle - \langle {\bf u}\cdot \nabla\times {\bf u}\rangle\right)\,dq, \label{appen_EDQNM_alpha} \\
\beta_{QN} &=& \frac{2}{3}\int^{t} \Theta_{kpq}(t)\langle u^2 \rangle\,dq + \frac{2}{3}\int^{t} \Theta_{kpq}(t)\langle b^2 \rangle\,dq, \label{appen_EDQNM_beta}
\end{eqnarray}
The relaxation time $\Theta_{kpq}$ is given by $\frac{1 - \exp(-\mu_{kpq}t)}{\mu_{kpq}}$, which converges to a constant over time: $\Theta_{kpq} \rightarrow \mu^{-1}_{kpq}$. The eddy damping operator $\mu_{kpq}$ is determined experimentally. Note that the coefficients $\alpha$ and $\beta$ have a factor of $2/3$, which stems from the quasi-normalization process that reduces fourth-order moments to second-order ones.\\

The EDQNM approach requires an additional time differentiation of the momentum and magnetic induction equations, resulting in fourth-order terms. In EDQNM, these fourth-order moments, $ \langle x_l x_m x_n x_q \rangle $, are approximated by second-order moments under the assumptions of isotropy and homogeneity, i.e., $ \sum_{lmnq} \langle x_l x_m \rangle \langle x_n x_q \rangle $. The second-order moments are then expressed in terms of $ E_V, E_M, H_V, H_M $, and the cross helicity $ \langle \mathbf{u} \cdot \mathbf{b} \rangle $. However, incorporating cross helicity into EDQNM requires more extensive theoretical calculations than the other terms, complicating the results. Typically, the EDQNM approach does not account for its effect, although it plays a role in decreasing the EMF $\langle \mathbf{u}\times {\bf b}\rangle$ dynamo effect.\\

\noindent (iv) Test field method \citep{2005AN....326..245S}:\\
If the simulation with the artificial test field $\overline{B}^T$ is repeated, data sets for $\mathbf{u}$ and $\mathbf{b}$ can be obtained. Then, from
\[
\boldsymbol{\xi}_i = \langle \mathbf{u} \times \mathbf{b} \rangle_i = \alpha_{ij} \overline{B}^T_j + \beta_{ijk} \frac{\partial \overline{B}^T_j}{\partial x_k} + \gamma_{ijlk} \dots
\]
the coefficients can be calculated. TFM provides detailed information on $\alpha_{ij}$ and $\beta_{ijk}$ depending on the component and position \citep{2013MNRAS.432.1651D, 2015ApJ...811..135A}. In particular, TFM reveals that the time-averaged magnetic diffusion effect, such as $ \beta_{r\theta} $ and $ \beta_{r\phi} $, is effectively negative, which is consistent with the negative $ \beta $ effect observed in our study \citep{2009A&A...500..633K, 2013MNRAS.432.1651D}. However, the validity of $ \overline{B}^T $ still requires further consideration. When the test magnetic field is applied, charged particles move freely along the direction of $ \overline{B}^T $, while their motion perpendicular to $ \overline{B}^T $ is constrained by the field. As $ \overline{B}^T $ increases, the Larmor radii of the particles shrink, and the electric Coulomb interactions, along with the binding energy between the particles, are altered. This leads to a particle distribution geometry resembling a needle shape. As a result, the distribution function $ f $ becomes anisotropic, requiring the continuity and momentum equations to be split into parallel and perpendicular components. The anisotropic nature introduced by the external $\overline{B}^T $ field cannot be disregarded. Of course, as assumed in TFM, if the size of the test field is very small, the impact on the distribution function will be limited. Nonetheless, it is still uncertain whether such a hypothetical field can be applied to actual observations.\\

\subsection{Statistical Second Order Moment}
In DIA and EDQNM, the second-order moments such as $\langle u_iu_j\rangle$ or $\langle b_ib_j\rangle$ needs to be replaced by energy and helicity to close the equations. These methods, along with mean-field theory (MFT), utilize two-point cross-correlation moments, such as those between velocity and magnetic field fluctuations, in contrast to quantities like kinetic energy ($E_V$) and kinetic helicity ($H_V$), which are based on single-point correlations. The corresponding statistical-mechanical justification is given below \citep{2008tufl.book.....L}.
\begin{eqnarray}
U_{jm}\equiv\langle u_j(r)u_m(r+l)\rangle=A(l)\delta_{jm}+B(l)l_jl_m+C(l)\epsilon_{jms}l_s.
\label{appen_general_beta_derivation1}
\end{eqnarray}
With $\vec{l}=(l,\,0,\,0)$ or other appropriate coordinates, we infer the relation of `$A$', `$B$', and `$C$' as follows: $A+l^2B\equiv F$, $A\equiv G$, $(U_{23}=)lC\equiv H$. Then, Eq.(\ref{appen_general_beta_derivation1}) is represented as
\begin{eqnarray}
U_{jm}=G\,\delta_{jm}+\frac{(F-G)}{l^2}\,l_jl_m+H\epsilon_{jms}\frac{l_s}{l}.
\label{appen_general_beta_derivation2}
\end{eqnarray}
With $\nabla \cdot {\bf U}=0$, we get the additional constraint.
\begin{eqnarray}
\frac{\partial U_{jm}}{\partial l_j}=\frac{l_j}{l}G'\delta_{jm}+4l_m\frac{F-G}{l^2}+l_m\frac{(F'-G')l^2-2l(F-G)}{l^3}=0,
\label{appen_constraint_F_G}
\end{eqnarray}
which leads to $G=F+(l/2)\,\partial F/\partial l$. So, the second order moment is
\begin{eqnarray}
U_{jm}=\bigg(F+\frac{l}{2}\frac{\partial F}{\partial l}\bigg)\,\delta_{jm}-\frac{l}{2l^2}\frac{\partial F}{\partial l}l_jl_m+H\epsilon_{jms}\frac{l_s}{l}.
\label{appen_general_beta_derivation3}
\end{eqnarray}
If $ j = m $, then $ U_{jj} = F = u^2/3 = E_V/6$. Conversely, if $ j \neq m $, the relation
\[
\left\langle \epsilon_{ijk} u_j(r) u_m(r+l) \frac{\partial \overline{B}_k}{\partial \overline{r}_m} \right\rangle \rightarrow -\left\langle \frac{\epsilon_{ijk} l_j l_m}{2l} \frac{\partial F}{\partial l} \right\rangle \frac{\partial \overline{B}_k}{\partial \overline{r}_m}
\]
indicates that any value of $ m $ renders the average negligible. For the term $ H $, we utilize Lesieur's approach \citep{2008tufl.book.....L}:
\begin{eqnarray}
H_V&=&\lim_{y\rightarrow x} \bf u(x)\cdot \nabla\times {\bf u(y)}\nonumber\\
&=&\lim_{y\rightarrow x}\epsilon_{ijn} u_i\frac{\partial u_n}{\partial y_j} =\lim_{l\rightarrow 0}\epsilon_{ijn} \frac{\partial U_{in}(l)}{\partial l_j}\,\,\,(\leftarrow y=x+l)\nonumber\\
&=&\lim_{l\rightarrow 0}\epsilon_{ijn} \epsilon_{ins}\bigg(\delta_{js}\frac{H}{l}-\frac{l_jl_s}{l^3}H+\frac{l_jl_s}{l^2}\frac{\partial H}{\partial l}\bigg)=-\frac{6}{l}H
\label{appen_general_beta_derivation4}
\end{eqnarray}

Then, $U_{jm}$ is
\begin{eqnarray}
&&\langle u_j(r)u_m(r+l)\rangle=\frac{2}{3}E_V\,\delta_{jm}-\epsilon_{jms}\frac{l_s}{6}H_V\mathrm,\\
&&\nonumber\\
&&\langle b_j(r)b_m(r+l)\rangle=\frac{2}{3}E_M\,\delta_{jm}-\epsilon_{jms}\frac{l_s}{6}H_C\mathrm,\\
\label{appen_general_beta_derivation5}
\end{eqnarray}
Here, the current helicity $H_C$ is defined as $H_C = \langle \mathbf{b} \cdot \nabla \times \mathbf{b} \rangle$. It is a pseudoscalar, resulting from the dot product between an axial vector $\mathbf{b}$ and a polar vector $\nabla \times \mathbf{b}$.\\

\textbf{Data Availability}\\
All data are incorporated into the article and its online supplementary material.

\section*{Acknowledgements}
The author acknowledges the support from the physics department at Soongsil University.

\bibliography{bibfile_2024_1030}
\bibliographystyle{aasjournal}

\end{document}